\documentclass[a4paper,cite,11pt,psnfss]{article}

\usepackage[T1]{fontenc}
\usepackage[latin1]{inputenc}
\usepackage{comment}
\usepackage{ulem}
\usepackage{adjustbox}
\usepackage{appendix}
\usepackage{framed,epsfig,pgfplots,amsmath}
\usepackage{tikz}
\usepackage{tikz-feynhand}
\usepackage{tikz-feynman}
\usetikzlibrary{snakes,arrows,shapes,positioning,automata,backgrounds,calc,er,patterns}
\usepackage{mathtools, amsmath, amsthm, amssymb,amsfonts}
\usepackage{makecell}
\usepackage{bm}

\usepackage[usenames,dvipsnames,tree]{pstricks}

\usepackage{graphicx,color,pst-plot}

\usepackage{latexsym,amsmath,amsfonts,amssymb,amstext,mathrsfs,upgreek}

\newcommand{\hyper}[5]{\;_{#1}{\rm F}_{#2} \left(\left.\begin{} {#3}
\\ {#4} \end{matrix}\right| #5\right) }

\newmuskip\pFqmuskip
\newcommand*\pFq[7][8]{%
  \begingroup 
  \pFqmuskip=#1mu\relax
  \mathchardef\normalcomma=\mathcode`,
  \mathcode`\,=\string"8000
  \begingroup\lccode`\~=`\,
  \lowercase{\endgroup\let~}\pFqcomma
  {}_{#2}{#3}_{#4}{\left[\genfrac..{0pt}{}{#5}{#6};#7\right]}%
  \endgroup
}

\newcommand{\pFqcomma}{{\normalcomma}\mskip\pFqmuskip}

\allowdisplaybreaks[1]

\begin{document}

\begin{titlepage}
\begin{flushright}
%
\end{flushright}
\begin{flushright}
\end{flushright}

\vfill

\begin{center}

{\Large\bf  Massive One-loop Conformal Feynman Integrals and\\}
\vspace{0.5cm}
{\Large\bf Quadratic Transformations of Multiple Hypergeometric Series\\}
 \vspace{1.5cm}
{\Large\bf }

\vfill

{\bf B. Ananthanarayan$^{a\ast}$, Sumit Banik$^{a\dagger}$, Samuel Friot$^{b,c\ddagger}$ and Shayan Ghosh$^{d\star}$}\\[1cm]
{$^a$ Centre for High Energy Physics, Indian Institute of Science, \\
Bangalore-560012, Karnataka, India}\\[0.5cm]
{$^b$ Universit\'e Paris-Saclay, CNRS/IN2P3, IJCLab, 91405 Orsay, France } \\[0.5cm]
{$^c$ Univ Lyon, Univ Claude Bernard Lyon 1, CNRS/IN2P3, \\
 IP2I Lyon, UMR 5822, F-69622, Villeurbanne, France}\\[0.5cm]
{$^d$ Helmholtz-Institut f\"ur Strahlen- und Kernphysik \& Bethe Center for Theoretical Physics, Universit\"at Bonn, D-53115 Bonn, Germany} \\
\end{center}
\vfill

\begin{abstract}

The computational technique of $N$-fold Mellin-Barnes (MB) integrals, presented in a companion paper by the same authors, is used to derive sets of series representations of the massive one-loop conformal 3-point Feynman integral in various configurations. This shows the great simplicity and efficiency of the method in nonresonant cases (generic propagator powers) as well as some of its subtleties in the resonant ones (for unit propagator powers). We confirm certain results in the physics and mathematics literature and provide many new results, some of them dealing with the more general massive one-loop conformal $n$-point case. In particular, we prove two recent conjectures that give the massive one-loop conformal $n$-point integral (for generic propagator powers) in terms of multiple hypergeometric series.  
We show how these conjectures, that were deduced from a Yangian bootstrap analysis, are related by a tower of new quadratic transformations in Hypergeometric Functions Theory.  Finally, we also use our MB method to identify spurious contributions that can arise in the Yangian approach.

\end{abstract}

\vspace{1cm}
\small{$\ast$ anant@iisc.ac.in }

\small{$\dagger$ sumitbanik@iisc.ac.in}

\small{$\ddagger$ samuel.friot@universite-paris-saclay.fr}

\small{$\star$ ghosh@hiskp.uni-bonn.de}

\end{titlepage}

\section{Introduction}

In \cite{ABFGgeneral}, we have presented the first systematic computational method allowing us to derive sets of series representations of $N$-fold Mellin-Barnes (MB) integrals, in the case where $N$ is a given positive integer. Our method was described at a general level and we have given only one simple twofold example as an application to provide a quick understanding of the method and of our notation. Here we show the details of a few more realistic applications, so that possible calculational subtleties can better be handled. This is one of the aims of this paper, where we apply our technique to the calculation of the massive one-loop conformal 3-point integral with generic propagator powers, as well as in the case where these propagator powers are equal to unity. Further, the massive one-loop conformal $n$-point integral will also be studied. Apart from showing explicit computational details, these examples will also give us the opportunity to present many new results which are relevant in both Quantum Field Theory (QFT) and Hypergeometric Functions Theory. 

The massive one-loop conformal 3-point integral has been recently studied in the Yangian bootstrap analysis of \cite{Loebbert:2020hxk,Loebbert:2020glj}. An analytic expression for the case of generic propagator powers has been given for the first time in \cite{Loebbert:2020hxk}, in terms of the well-known Srivastava's $H_C$ triple series \cite{Srivastava67, Srivastava}. In \cite{Loebbert:2020glj}, an alternative expression, still in terms of a single triple hypergeometric series, has been derived by choosing another set of conformal variables. Each of these two results has contributed to guide the authors of \cite{Loebbert:2020glj} to conjectures for the expressions of the massive one-loop conformal $n$-point integral with generic propagator powers in two kinematic regions called A and B. The unit propagator powers case of the 3-point integral has also been considered for both regions in \cite{Loebbert:2020glj}, where a comparison with some results of the literature is given.

In the present work, starting from the two MB representations (associated with the two choices of conformal variables alluded to above) of the massive one-loop conformal $n$-point integral, we first begin by giving the proofs of the two conjectures of \cite{Loebbert:2020glj} which have been named Conjecture in Region A (respectively B). 

We later focus on the $n=3$ case and we show how to derive, apart from the Yangian bootstrap results of \cite{Loebbert:2020hxk,Loebbert:2020glj}, sets of series representations for the 3-point integral which are analytic continuations of one another, converging in various regions of the conformal variables space. Each set provides 13 additional series representations. Some of the analytic continuation formulas associated with the result in region A, for the generic propagator powers case, have been known for a long time due to prior studies of the $H_C$ triple hypergeometric series \cite{Srivastava72, Srivastava}. However, to our knowledge, apart from the latter and the four results given in \cite{Loebbert:2020glj}, the other formulas that we present in this paper are new. 

We underline here that, in addition to these interesting results, our MB approach \cite{ABFGgeneral} is used in Section \ref{Set1series} to show how some of the zeros derived in the Yangian bootstrap approach \cite{Loebbert:2020hxk,Loebbert:2020glj} are spurious and, thus, should be removed from the sets of solutions corresponding to several cases studied in these papers. This appears to be a drawback of the Yangian approach which happens when the object under study has an MB representation which has gamma functions in the denominator of its integrand. 

Other distinctive features have been noted, that distinguish the efficiency of the Yangian bootstrap and our MB method. An additional numerical analysis is needed in the Yangian approach in order to determine the precise form of the overall coefficients of the series representations whereas this is not the case in our approach where the full expressions can be computed directly from the MB integral representation. Furthermore, in the case where the series representation of a given Feynman integral is a linear combination of several series, this combination cannot be derived in the Yangian approach without the use of some shift identities, which are not guaranteed to exist, or without the knowledge of the convergence properties of each of the series solutions obtained from this approach \cite{Loebbert:2019vcj}, which can be very large in nontrivial examples (for instance as large as 2530 in the case of the conformal hexagon \cite{Loebbert:2019vcj,Ananthanarayan:2020ncn}). This necessary external input can thus be extremely difficult to find.

In contrast to the Yangian approach, our MB method \cite{ABFGgeneral} does not require any convergence consideration to determine these linear combinations. This does not mean that convergence issues are uninteresting and, in fact, once the series representations have been derived, it may even be necessary to know where they converge (when they do \cite{ABFGgeneral}). Once again our approach has, in many cases, a great advantage at reducing the study of the possibly many series that constitute a given series representation, to just a single one: the master series \cite{ABFGgeneral}. It just so happens that the study of the master series as well as the convergence properties of the series representations that we obtained from our general MB method for the massive one-loop conformal 3-point integral put us on the road to find several further interesting results. Indeed, as explained in Section \ref{quadratic}, after having proved that for the 3-point integral, region B is included in region A, we were naturally led to the discovery of a new quadratic transformation formula for $H_C$ in terms of the other well-known Srivastava $H_B$ triple hypergeometric series \cite{Srivastava64,Srivastava}. We inferred from this result that the conjectured expressions of \cite{Loebbert:2020glj} for the $n$-point integral in region A and B can be interpreted as the LHS and the RHS of a tower of quadratic transformation formulas for hypergeometric functions of $n(n-1)/2$ variables which, apart from the lowest-order $n=2$ case, which involves the $_2F_1$ Gauss hypergeometric series, seem to be unknown in hypergeometric functions theory. 
This comes from the fact that at one loop the conformal constraint, that fixes the sum of the generic propagator powers to be equal to the spacetime dimension, has the same form as the constraint satisfied by the parameters of hypergeometric functions obeying certain quadratic transformations. 
We conclude that, since these quadratic transformations are the consequence of an appropriate choice of the conformal variables in the Feynman integral, this is an indication that QFT can be used as a tool to derive new results in hypergeometric functions theory. 

Another result that can be obtained due to this cross-fertilization between QFT and hypergeometric functions theory is the following.
One of the series representations that we obtained for the 3-point integral, analytically continuing the $H_C$ series result of region A, has a convergence region which is harder to compute than the other series representations and for which we only give a conjectured expression in this paper, as well as a proof of the maximal region that it can fill.
However, in the set of analytic continuations of the result of region B, obtained due to the alternative choice of conformal variables, it is possible to find a series representation whose convergence region includes this maximal region. This shows once again that QFT helps by providing a way to derive an alternative expression which is an analytic continuation of $H_C$ in this particular region, easier to handle and with a bigger convergence region than the one directly derived from the MB representation of $H_C$ itself. The same formula can also be used to provide another (new) quadratic transformation formula for $H_C$.

In fact, even better than this, the convergence regions of all the series representations that we have obtained as analytic continuations of the result of region B are wider than those analytically continuing the result of region A (although the former do not include the latter in general). Therefore, the set of analytic continuations of region B's result considerably reduces, in the present case of the massive one-loop conformal 3-point integral, the domain of the conformal variables space which does not belong to any of the convergence regions of the set of series representations associated with region A's result (this particular "unreachable" domain, that we have studied and called \textit{white region} in some previous works \cite{Ananthanarayan:2019icl,Ananthanarayan:2020acj,Ananthanarayan:2020xut}, is inherent to the series representations of many MB integrals).

The plan of this paper is as follows: in Section \ref{Proof_conj} we give a brief recapitulation of the method introduced
in \cite{ABFGgeneral} and then present proofs of the $n$-point conjectures. In Section \ref{MB_3point} we
present MB representations of the 3-point integral and the nonresonant case of generic propagator powers is discussed in Section \ref{Non_Resonant}.
In Section \ref{quadratic} we describe a new tower of quadratic transformations for a certain class of multiple hypergeometric series, that arise from the QFT analysis.
In Section \ref{resonant} the analysis of Section \ref{Non_Resonant} is performed for the resonant
case with unit propagator powers. In Section \ref{conclusions} we present our conclusions. An Appendix brings together all relevant results not explicitly stated in the main core of the text.

\section{The Method and Proofs of the $n$-point Conjectures in Regions A and B\label{Proof_conj}}

In this section we give a short summary, for the nonresonant case, of the MB computational technique introduced in \cite{ABFGgeneral} and we apply it on two different MB representations of the massive dual-conformal $n$-point one-loop Feynman integral to prove the conjectures proposed in \cite{Loebbert:2020glj}.

There are two conjectures giving the expression of the $n$-point integral in two kinematical regions A and B. Each of these regions corresponds to the choice of a particular set of conformal variables. We call these sets Set 1 and Set 2 in the following.

It is easy to prove the conjectures from the MB representations of the $n$-point function obtained by choosing one set or the other as we show later in this section. Furthermore, we will also state in Section \ref{quadratic} that these two conjectures are nothing but the LHS and RHS of a tower of quadratic transformation formulas which, to our knowledge, are new results in hypergeometric functions theory.

Some of the conventions, notation and terminology in the following sections are taken from \cite{ABFGgeneral} and \cite{Loebbert:2020glj}.

\subsection{The Conic Hull method: a brief introduction\label{remainder}
}
A general $N$-fold MB representation is of the form:
\begin{align} \label{N_MB}
    I &(x_1,x_2,\cdots ,x_N) = \int\limits_{-i \infty}^{+i \infty} \frac{ \text{d} z_1}{2 \pi i} \cdots \int\limits_{-i \infty}^{+i \infty}\frac{ \text{d} z_N}{2 \pi i}\,\,  \frac{\prod\limits_{i=1}^{k} \Gamma^{a_i}({\bf e}_i\cdot{\bf z}+g_i)}{\prod\limits_{j=1}^{l} \Gamma^{b_j}({\bf f}_j\cdot{\bf z}+h_j)} x^{z_1}_{1} \cdots x^{z_N}_{N}
\end{align}
where $a_i , b_j, k, l$, $N$ are positive integers, ${\bf z}=(z_1, \cdots, z_N)$, ${\bf e}_i$ and ${\bf f}_j$ are $N$-dimensional real vectors, $g_i$ and $h_i$ are reals, while the variables $x_1 , \cdots , x_N$ can be complex. 
The integration contours separate the sets of poles of the gamma functions in the numerator of the MB integrand in the usual way \cite{Smirnov:2012gma}.

We look for multiple series representations of Eq.(\ref{N_MB}). The type of these series strongly depends on the vector ${\bf \Delta}=\sum_{i=1}^{k} {\bf e}_i - \sum_{j=1}^{l} {\bf f}_j$ \cite{ABFGgeneral,Passare:1996db,TZ}. Indeed, one can classify the degenerate and nondegenerate cases by the condition ${\bf \Delta}=\vec{0}$ and ${\bf\Delta} \neq \vec{0}$ (where, in the latter case, none of the ${\bf e}_i$ is proportional to ${\bf \Delta}$), respectively. In the degenerate case, one can obtain several convergent series representations of Eq.(\ref{N_MB}), which converge in various regions of the ${\bf x}=(x_1 , \cdots , x_N)$ parameter space, while in the nondegenerate situation, only one series representation converges, the others having a diverging asymptotic behavior.

To obtain the series representations of Eq.\eqref{N_MB}  using the conic hull theory \cite{ABFGgeneral} for the simple nonresonant case, where at all poles exactly $N$ singular hyperplanes associated with the gamma functions of the numerator of the integrand in Eq.(\ref{N_MB}) intersect, we proceed as follows \cite{ABFGgeneral}:
\begin{itemize}
    \item One considers all possible $N$-combinations of gamma functions in the numerator of the MB integrand and retains only those combinations for which the associated matrix
    \begin{align}
        A=( {\bf e}_{i_1}, \cdots , {\bf e}_{i_N})^{T}
    \end{align}
    is nonsingular, where ${\bf e}_{i_1}, \cdots, {\bf e}_{i_N}$ are the coefficient vectors of the gamma functions in the combination.
    \item One then assigns a series to each retained $N$-combination, which we call a \textit{building block} and denote by $B_{i_1,\cdots,i_N}$. The series is obtained by summing over the residues of only those poles which originate from the intersection of singular hyperplanes of the gamma functions in the $N$-combination, and by dividing by $|\text{det}(A)|$.
    \item One further assigns a conic hull to each building block $B_{i_1,\cdots,i_N}$, whose vertex is at the origin and edges along the vectors ${\bf e}_{i_1}, \cdots, {\bf e}_{i_N}$. Therefore, the parametric representation of the conic hull is
    \begin{align}
      \{s_1 {\bf e}_{i_1} + \cdots + s_N {\bf e}_{i_N} | s_j \in \mathbb{R}^+, (j=1,...,N) \}
    \end{align}
    \item Finally, the series representations of the MB in Eq.(\ref{N_MB}) are obtained by summing the building blocks associated with the conic hulls that form the largest subsets, in the set of all conic hulls, having a nonempty intersection region.
\end{itemize}
We have also introduced in \cite{ABFGgeneral} the notion of \textit{master series} for each representation of a degenerate MB integral, which can be obtained by mapping back the corresponding common intersection conic hull (master conic hull) of a series representation of the MB integral, to a series, in the opposite manner as a building block is mapped to a conic hull. We conjecture that the convergence region of the master series and of the corresponding series representation of the MB integral will be the same if there is no cancellation of poles (or only a finite number of cancellations) by the denominator in the MB integrand. More technical details on the nonresonant case and the procedure to treat the resonant case, where at some poles more than $N$ singular hyperplanes intersect, can be found in \cite{ABFGgeneral}.

\subsection{The massive $n$-point conformal Feynman integral at one-loop}
\begin{figure}[h]
\begin{center}
\includegraphics[scale=1]{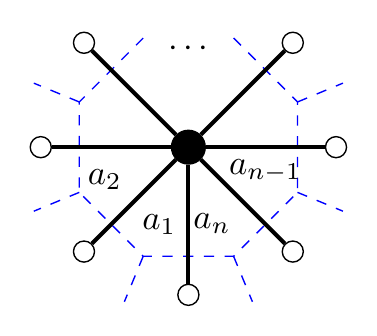}
\end{center}
\caption{Massive dual-conformal $n$-point one-loop Feynman integral.\label{OneLoopNpoint}}
\end{figure}

The dual-conformal $n$-point one-loop massive Feynman integral (see Fig.\ref{OneLoopNpoint}) has the following integral representation \cite{Loebbert:2020glj}
\begin{align}\label{N-point-integral}
    I_{n \bullet}^{m_{1} \ldots m_{n}}= \int \frac{\text{d}^{D}x_{0}}{\prod_{i=1}^{n} \left( x^{2}_{0 i} + m^{2}_i\right)^{a_{i}}}
\end{align}
where $x^{\mu}_{ij}=x^{\mu}_i-x^{\mu}_j$ and the dualized momenta $x_i$ are defined as $p^{\mu}_i=x^{\mu}_{i}-x^{\mu}_{i+1}$. The mass and power of the $i^{\text{th}}$ propagator is denoted by $m_i$ and $a_i$ respectively, which take generic nonzero values. Due to conformal symmetry, the following constraint is imposed $\sum_{i=1}^{n} a_i=D$ on the propagator powers, where $D$ is the spacetime dimension.

\subsection{Conformal variables Set 1\label{conf_set1} (Conjecture in Region A)}

Let us take the set of conformal variables that the authors of \cite{Loebbert:2020glj} have chosen for what they call region A, which reads $u_{ij}=\frac{x_{ij}^2+(m_i-m_j)^2}{-4m_im_j}$.

In terms of these, the $n$-point integral in Eq.\eqref{N-point-integral} has the following Feynman parameterization \cite{Loebbert:2020glj}
\begin{align}\label{FPrep}
&I_{n \bullet}^{m_{1} \ldots m_{n}}=\nonumber \\& \ \ \ \pi^{\frac{D}{2}}\Gamma\left(\frac{D}{2}\right)\left(\prod_{i=2}^n\int_0^\infty\text{d}\alpha_i\right)\left(\prod_{i=1}^n\frac{\alpha_i^{a_i-1}}{\Gamma(a_i)m_i^{a_i}}\right)\left.\left(\sum_{i<j=1}^n2\alpha_i\alpha_j(1-2u_{ij})+\sum_{i=1}^n\alpha_i^2\right)^{-\frac{D}{2}}\right\vert_{\alpha_1=1}
\end{align}

From this formula, we can derive the $\frac{n(n-1)}{2}$-fold MB representation
\begin{align}
     I_{n \bullet}^{m_{1} \ldots m_{n}}= \frac{\pi^{D/2+1/2}}{2^{D-1}\prod_{i=1}^{n}  \Gamma(a_i)m_{i}^{a_{i}}} \prod_{\alpha\in B_{n} } \left( \int_{-i\infty}^{+i\infty} \frac{\text{d}z_{\alpha}}{2 \pi i} \Gamma(-z_{\alpha})(-u_{\alpha})^{z_\alpha} \right) \frac{\prod_{i=1}^{n}\Gamma\left(a_i+\sum_{\alpha \in B_{n|i}} z_{\alpha}\right)}{\Gamma\left( \frac{D+1}{2}+\sum_{\alpha \in B_{n}} z_{\alpha}\right)}
     \label{MB}
\end{align}
where $B_n=\{12,13,23,\cdots,(n-1,n)\}$ is the set of pairs of distinct integers (written in increasing order) in $\{1,\cdots,n\}$ and $B_{n\vert j}$ is the subset of $B_n$ with pairs containing $j$. The integration contours separate the sets of poles of the gamma functions in the numerator of the MB integrand in the usual way \cite{Smirnov:2012gma}.

One first notes that the $\frac{n(n-1)}{2}$-dimensional vector ${\bf \Delta}$, as defined in Section \ref{remainder}, is null which means that the MB integral in Eq.(\ref{MB}) belongs to the degenerate class \cite{ABFGgeneral,Passare:1996db,TZ}. Therefore, several series representations of this integral coexist, converging in different regions of the $(u_{12},u_{13},...,u_{n-1,n})$-space. Moreover, since the powers $a_i$ of the propagators of the $n$-point integral are generic, it is a nonresonant case.
	
Now, using our MB computational method \cite{ABFGgeneral}, it is obvious from the form of the MB representation in Eq.(\ref{MB}) that the (trivial) conic hull associated with the $\frac{n(n-1)}{2}$-combination made of the first gamma functions $\Gamma(-z_{\alpha})$ $(\alpha=12,13,\cdots,(n,n-1))$ in the numerator of the MB integrand belongs to the $(-,-,...,-)$ negative hyperquadrant of the $(z_{12},z_{13},...,z_{n-1,n})$-space. Therefore, since the normal vectors associated with all the other gamma functions of the numerator are in the $(+,+,...,+)$ positive hyperquadrant, it is impossible for any conic hull formed by a set of normal vectors that possess at least one of these positive normal vectors (\textit{i.e} for all other conic hulls associated to this MB integral) to have a nonempty intersection with the trivial conic. 
Hence, we conclude that there exists a series representation of the MB integral in Eq.(\ref{MB}) that consists in a single series, associated with the trivial conic hull and whose sets of singular points are located at $(-N_{12},-N_{13},...,-N_{n-1,n})$. 

To compute this series representation following \cite{ABFGgeneral}, we first perform the change of variable $z_\alpha\rightarrow z_\alpha+N_\alpha$ so as to bring these singularities to the origin. Then we explicitly extract the singular factors of the gamma functions that are singular at the origin, by an application of the generalized reflection formula $\Gamma(z-m)=\frac{\Gamma(1+z)\Gamma(1-z)(-1)^m}{z\ \Gamma(m+1-z)}$, $m\in \mathbb{Z}$. Omitting the overall factor, this gives to the MB integrand the form
\begin{align}
     \prod_{\alpha\in B_{n} } \left( \frac{1}{2 \pi i} \frac{\Gamma(1-z_{\alpha})\Gamma(1+z_{\alpha})(-1)^{N_\alpha}}{(-z_\alpha)\Gamma(N_\alpha+1+z_{\alpha})}(-u_{\alpha})^{z_\alpha+N_\alpha} \right) \frac{\prod_{i=1}^{n}\Gamma\left(a_i+\sum_{\alpha \in B_{n|i}}(z_{\alpha}+N_\alpha)\right)}{\Gamma\left( \frac{D+1}{2}+\sum_{\alpha \in B_{n}} (z_{\alpha}+N_\alpha)\right)}
     \label{MBint}
\end{align}
Now, the calculation of the corresponding Cauchy residue can be achieved, as indicated in \cite{ABFGgeneral}, by simply dividing the above expression by $\vert \text{det}A\vert$ which is unity here as the $\frac{n(n-1)}{2}\times\frac{n(n-1)}{2}$ matrix $A=({\bf e}_1,{\bf e}_2, \cdots, {\bf e}_{n(n-1)/2})^T=((-1,0,\cdots,0),(0,-1,\cdots,0),\cdots,(0,0,\cdots,-1))^T$, by removing all the singular factors $-z_{12},-z_{13},\cdots,-z_{n-1,n}$ from the denominator and by putting  $z_{12}=z_{13}=\cdots=z_{n-1,n}=0$. One then obtains
\begin{align}
 \text{Res.}=  \prod_{\alpha\in B_{n} } \left( \frac{1}{2 \pi i} \frac{(-1)^{N_\alpha}}{\Gamma(N_\alpha+1)}(-u_{\alpha})^{N_\alpha} \right) \frac{\prod_{i=1}^{n}\Gamma\left(a_i+\sum_{\alpha \in B_{n|i}}N_\alpha\right)}{\Gamma\left( \frac{D+1}{2}+\sum_{\alpha \in B_{n}}N_\alpha\right)}
     \label{MBint_res}
\end{align}
Summing over all residues and including the overall factor, one then finally finds
\begin{align}
 I_{n \bullet}^{m_{1} \ldots m_{n}}=  \frac{\pi^{D/2+1/2}}{2^{D-1}\prod_{i=1}^{n}  \Gamma(a_i)m_{i}^{a_{i}}}  \sum_{N_{12},N_{13},\cdots,N_{n-1,n}=0}^\infty\prod_{\alpha\in B_{n} } \left(\frac{u_{\alpha}^{N_\alpha}}{N_\alpha!} \right) \frac{\prod_{i=1}^{n}\Gamma\left(a_i+\sum_{\alpha \in B_{n|i}}N_\alpha\right)}{\Gamma\left( \frac{D+1}{2}+\sum_{\alpha \in B_{n}}N_\alpha\right)}
     \label{MBint_res_final}
\end{align}
which, by using Pochhammer's symbols and the duplication formula of the gamma function, proves the conjecture of \cite{Loebbert:2020glj} in region A:
\begin{align}
 I_{n \bullet}^{m_{1} \ldots m_{n}}=  \frac{\pi^{D/2}\Gamma(D/2)}{\Gamma(D)\prod_{i=1}^{n} m_{i}^{a_{i}}}  \sum_{N_{12},N_{13},\cdots,N_{n-1,n}=0}^\infty  \frac{\prod_{i=1}^{n}\left(a_i\right)_{\sum_{\alpha \in B_{n|i}}N_\alpha}}{\left( \frac{D+1}{2}\right)_{\sum_{\alpha \in B_{n}}N_\alpha}}\prod_{\alpha\in B_{n} } \left(\frac{u_{\alpha}^{N_\alpha}}{N_\alpha!}\right)
     \label{conjecture_regA}
\end{align}

\subsection{Conformal variables Set 2 (Conjecture in Region B)}
Here the conformal variables are $v_{ij}=\frac{x_{ij}^2+m_i^2+m_j^2}{2m_im_j}$ (corresponding to region B \cite{Loebbert:2020glj}).
In this case, we replace $(1-2 u_{ij})$ by $v_{ij}$ in the Feynman parametrization in Eq.(\ref{FPrep}) to get the following MB representation:
\begin{align}
     I_{n \bullet}^{m_{1} \ldots m_{n}}= \frac{\pi^{D/2}}{2^{n-1}\prod_{i=1}^{n}  \Gamma(a_i)m_{i}^{a_{i}}} \prod_{\alpha\in B_{n} } \left( \int_{-i\infty}^{+i\infty} \frac{\text{d}z_{\alpha}}{2 \pi i} \Gamma(-z_{\alpha})(2v_{\alpha})^{z_\alpha} \right) \prod_{i=1}^{n}\Gamma\left(\hat{a}_i+\sum_{\alpha \in B_{n|i}} \hat{z}_{\alpha}\right)\label{MB2}
\end{align}
where $\hat a_i\doteq a_i/2$.

A similar reasoning as in Section \ref{conf_set1} gives the proof of the conjecture of \cite{Loebbert:2020glj} in region B which reads
\begin{align}
 I_{n \bullet}^{m_{1} \ldots m_{n}}=  \frac{\pi^{D/2}}{2^{n-1}} \frac{1}{\prod_{i=1}^{n}  \Gamma(a_i)m_{i}^{a_{i}}} \sum_{N_{12}, N_{13},\cdots,N_{n-1,n}=0}^\infty  \prod_{i=1}^{n}\Gamma\left(a_i/2+\sum_{\alpha \in B_{n|i}}N_\alpha/2\right)\prod_{\alpha\in B_{n} } \frac{\left(-2v_{\alpha}\right)^{N_\alpha}}{N_\alpha!}
     \label{conjecture_regB}
\end{align}

\section{MB representations of the massive one-loop conformal 3-point Feynman integral\label{MB_3point}}

\begin{figure}[h]
\begin{center}
\includegraphics[scale=1.1]{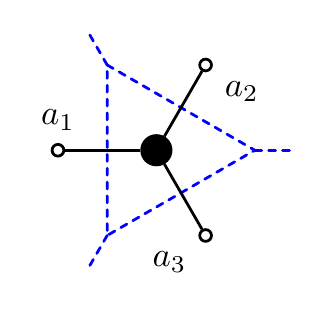}
\end{center}
\caption{Massive dual-conformal $3$-point one-loop Feynman integral.\label{OneLoop3point}}
\end{figure}

In this section, we focus on the $n=3$ case of Eqs.(\ref{MB}) and (\ref{MB2}) (see Fig.\ref{OneLoop3point}).

After giving the explicit form of these threefold MB integrals we will present a few general facts about them. We will then detail, in subsequent sections, the application of our systematic computational method of multiple MB integrals \cite{ABFGgeneral} on these simple albeit nontrivial threefold examples to show how the nonresonant cases can be easily handled with our technique and, afterward, to emphasize which subtle points one has to be careful of, in the resonant situations.

These calculations will give birth to many new formulas which, to our knowledge, are unknown in hypergeometric functions theory. In particular, we will show in Section \ref{quadratic} how their study lead us to the derivation of a tower of new quadratic transformation formulas, obtained from the conjectures that we have proved in Section \ref{Proof_conj}.

\subsection{Mellin-Barnes representation for Set 1}

Starting from Eq.(\ref{MB}) with $n=3$, we obtain the threefold MB representation of the massive one-loop conformal 3-point integral as
\begin{equation}
\begin{multlined}
    I_{3 \bullet}^{m_{1} m_2 m_{3}}=\frac{\pi^{D/2+1/2}}{2^{D-1}m_1^{a_1}m_2^{a_2}m_3^{a_3}\Gamma(a_1)\Gamma(a_2)\Gamma(a_3)}J(a_1,a_2,a_3; u,v,w)
    \end{multlined}
    \label{overall}
\end{equation}
where,
\begin{equation}
\begin{multlined}\label{MBtriangle}
     {J_1}(a_1,a_2,a_3; u,v,w)=\int\limits_{-i \infty}^{+i \infty} \frac{\text{d} z_{1}}{2 \pi i} \int\limits_{-i \infty}^{+i \infty} \frac{\text{d} z_{2}}{2 \pi i}\int\limits_{-i \infty}^{+i \infty} \frac{\text{d} z_{3}}{2 \pi i} \left(-u\right)^{z_{1}} \left(-v\right)^{z_{2}}  \left(-w\right)^{z_{3}} \Gamma (-z_{1})\Gamma(-z_{2})\Gamma(-z_{3}) \\ \times
    \frac{{\Gamma}(a_1+z_{1}+z_{2}){\Gamma}(a_2+z_{1}+z_{3}){\Gamma}(a_3+z_{2}+z_{3})}{\Gamma(D/2+1/2+z_1+z_2+z_3)}
    \end{multlined}
\end{equation}
and $u$, $v$, $w$ are conformal variables defined, respectively, as 
\begin{align}
    u &= \frac{x_{12}^{2}+(m_1-m_2)^{2}}{-4 m_1 m_2} \nonumber\\ 
    v &= \frac{x_{13}^{2}+(m_1-m_3)^{2}}{-4 m_1 m_3}\label{set1}\\ 
    w &= \frac{x_{23}^{2}+(m_2-m_3)^{2}}{-4 m_2 m_3}\nonumber
\end{align}\label{New_conformal_variables}
$D=a_1+a_2+a_3$ {,} is the conformal constraint.

The MB integration contours in Eq.\eqref{MBtriangle} separate the sets of poles of the gamma functions in the numerator of the MB integrand in the usual way  \cite{Smirnov:2012gma}.

Any experienced practitioner of MB integrals and hypergeometric functions will recognize Eq.(\ref{MBtriangle}), modulo an overall factor, as the MB representation of the well-known Srivastava's $H_C$ triple hypergeometric function \cite{Srivastava67, Srivastava}.

As we have already seen in Section \ref{Proof_conj}, for this MB integral ${\bf\Delta}=(0,0,0)$; therefore, it belongs to the degenerate class \cite{ABFGgeneral,Passare:1996db,TZ}. Thus, apart from the trivial $H_C$ representation, several other series representations coexist for this integral, which converge in different kinematic regions.

These series representations are analytic continuations of one another since the quantity
\begin{align}\label{Alpha_definition}
    \alpha= \text{Min}\,( |y_1|+|y_2|+|y_3|+|y_1+y_2|+|y_1+y_3|+|y_2+y_3|-|y_1+y_2+y_3|)
\end{align}
subject to the condition
\begin{align}
     y_1^2+y_2^2+y_3^2=1,
\end{align}
where the $y_i\ (i=1,2,3)$ are real numbers, is positive \cite{ABFGgeneral,Passare:1996db,TZ}.

This can be easily shown by applying the triangle inequality to the first three terms of Eq.\eqref{Alpha_definition} to obtain
\begin{align}
    \alpha \geq \text{Min}\,(|y_1+y_2|+|y_1+y_3|+|y_2+y_3|)
\end{align}
which is manifestly positive.
A minimization procedure indeed gives $\alpha=2$.

\subsection{MB representation for Set 2}

An alternative MB representation of the massive one-loop conformal 3-point integral, obtained from a second choice of conformal variables given by
\begin{align} 
    u' &= \frac{x_{12}^{2}+m_1^2+m_2^2}{2 m_1 m_2}=1-2u \nonumber\\ 
    v' &= \frac{x_{13}^{2}+m_1^2+m_3^2}{2 m_1 m_3}=1-2v\label{set2}\\ 
    w' &= \frac{x_{23}^{2}+m_2^2+m_3^2}{2 m_2 m_3}=1-2w\nonumber
\end{align}
is
\begin{align}
    I_{3 \bullet}^{m_{1} m_2 m_{3}}=\frac{\pi^{D/2}}{4\,m_1^{a_1}m_2^{a_2}m_3^{a_3}\Gamma(a_1)\Gamma(a_2)\Gamma(a_3)}J_{2}(a_1,a_2,a_3; u',v',w')\label{overall2}
\end{align}
where
\begin{align}\label{Conformal_Triangle_2}
   & J_2(a_1,a_2,a_3; u',v',w') = \int\limits_{-i \infty}^{+i \infty} \frac{\text{d} z_{1}}{2 \pi i} \int\limits_{-i \infty}^{+i \infty} \frac{\text{d} z_{2}}{2 \pi i}\int\limits_{-i \infty}^{+i \infty} \frac{\text{d} z_{3}}{2 \pi i} \left(2-4u\right)^{z_{1}} \left(2-4v\right)^{z_{2}}  \left(2-4w\right)^{z_{3}} \Gamma (-z_{1})\nonumber  \\  
   & \hspace{2cm}\times
    \Gamma(-z_{2})\Gamma(-z_{3})\Gamma\left(\frac{a_1+z_{1}+z_{2}}{2}\right)\Gamma\left(\frac{a_2+z_{1}+z_{3}}{2}\right)\Gamma\left(\frac{a_3+z_{2}+z_{3}}{2}\right)
\end{align}
and $u$, $v$, $w$ are the same conformal variables as defined in Eq.(\ref{set1}).

As in the case of Set 1, ${\bf \Delta}=(0,0,0)$ and $\alpha>0$.

\section{The nonresonant case: Generic propagator powers\label{Non_Resonant}}

For generic propagator powers $a_i\  (i=1,2,3)$, the MB representations of the 3-point integral given in Eqs.(\ref{overall}) and (\ref{overall2}) belong to the simple nonresonant class. It is therefore straightforward to derive their different series representations using our method \cite{ABFGgeneral}.
In this section, we obtain all the series representations in powers of the $u_\alpha$ and/or $\frac{1}{u_\alpha}$ that are analytic continuations of region A's result (Set 1) given in Eq.(\ref{MBint_res_final}), for $n=3$. The same approach will be followed for the second set of conformal variables and we will give the corresponding analytic continuations of region B's result (Set 2). Part of the formulas will be relegated to the Appendix. 

A similar analysis will show how to treat the resonant unit propagator powers case, on the example of Set 1 of conformal variables in Section \ref{resonant}. 

\subsection{Set 1: Series Representations\label{Set1series}}

To derive the series representations of the MB integral of Eq.(\ref{overall}), one first needs to obtain the set $S$ of building blocks from which the series representations will be composed. As the MB integral under study is threefold, this requires to determine all relevant 3-combinations of gamma functions in the numerator of the MB integrand, as well as their associated set $S'$ of conic hulls. 

As we have done in a similar exercise for a simpler twofold case in \cite{ABFGgeneral}, let us keep track of these gamma functions by labeling them by an integer $i$, as shown in Table \ref{Table1}, where we have tabulated the gamma functions along with the normal vectors and singular factors associated with them. 
\begin{table}[h]
\begin{center}
 \begin{tabular}{||c c c c||}
 \hline
 $i$ & $\Gamma$ function & ${\bf e}_i$ & ${\bf e}_i\cdot{\bf z}$ \\ [0.5ex] 
 \hline\hline
 1 & $\Gamma(-z_1)$ & $(-1,0,0)$ & $-z_1$ \\ 
 2 & $\Gamma(-z_2)$ & $(0,-1,0)$ & $-z_2$ \\
 3 & $\Gamma(-z_3)$ & $(0,0,-1)$ & $-z_3$ \\
 4 & ${\Gamma}(a_1+z_{1}+z_{2})$ &
 $(1,1,0)$ & $z_{1}+z_{2}$ \\
 5 & ${\Gamma}(a_2+z_{1}+z_{3})$ & $(1,0,1)$ & $z_{1}+z_{3}$ \\
 6 & ${\Gamma}(a_3+z_{2}+z_{3})$ & $(0,1,1)$ & $z_{2}+z_{3}$ \\[1ex]
 \hline
 \end{tabular}
 \end{center}
 \caption{ List of gamma functions in the numerator of the integrand in Eq.\eqref{MBtriangle} and their associated labels, normal vectors and singular factors.\label{Table1}}
 \end{table}
 
Since there are 6 gamma functions in the numerator of \eqref{MBtriangle}, one has $\binom{6}{3}=20$ possible choices of 3-combinations. Each choice is denoted by $(i_1,i_2,i_3)$, where $i_1$, $i_2$ and $i_3$ are the labels of the corresponding gamma functions given in the first column of Table 1.

\begin{figure}[htbp]
\centering
\includegraphics[width=6cm]{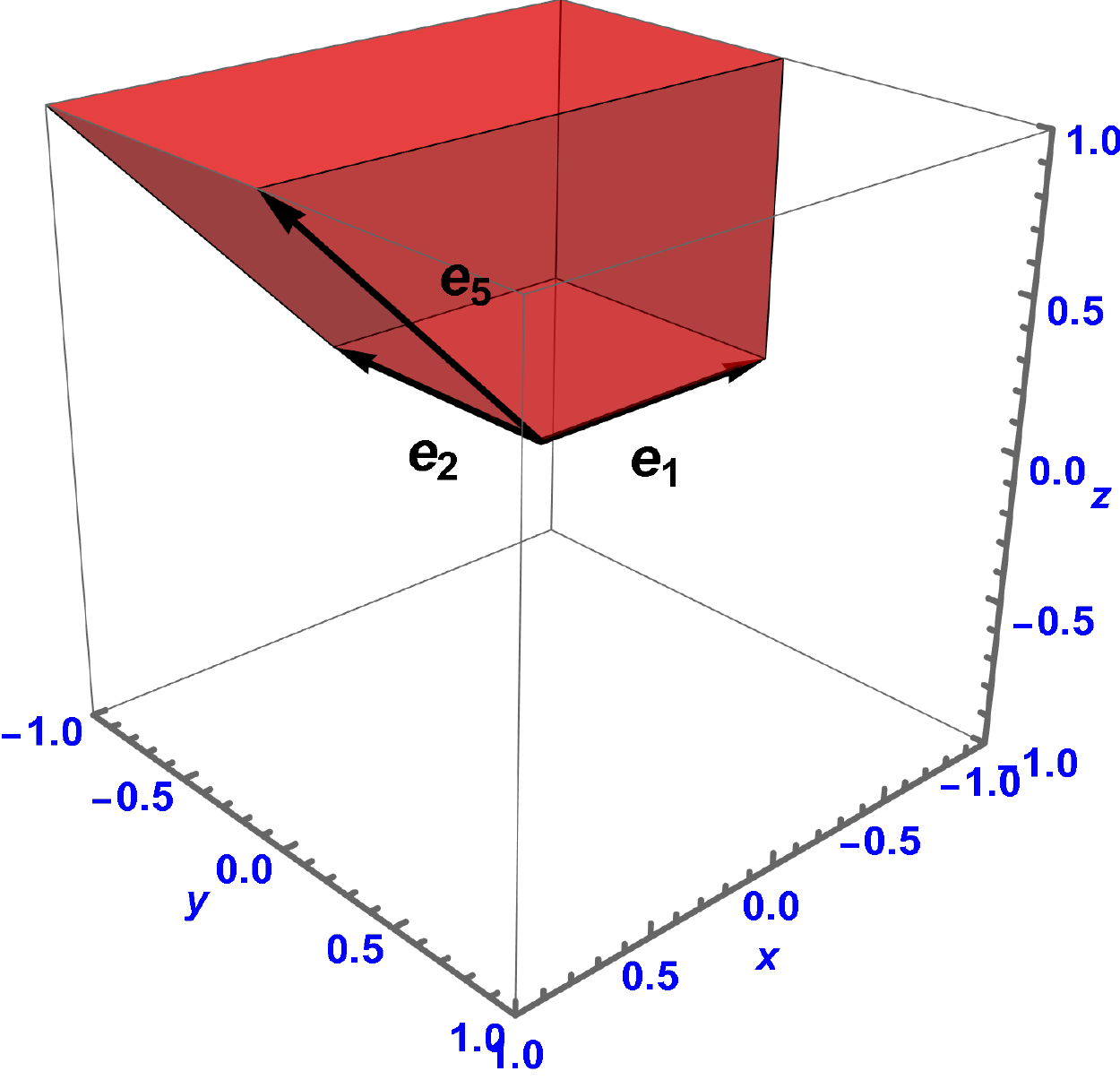}
\includegraphics[width=6cm]{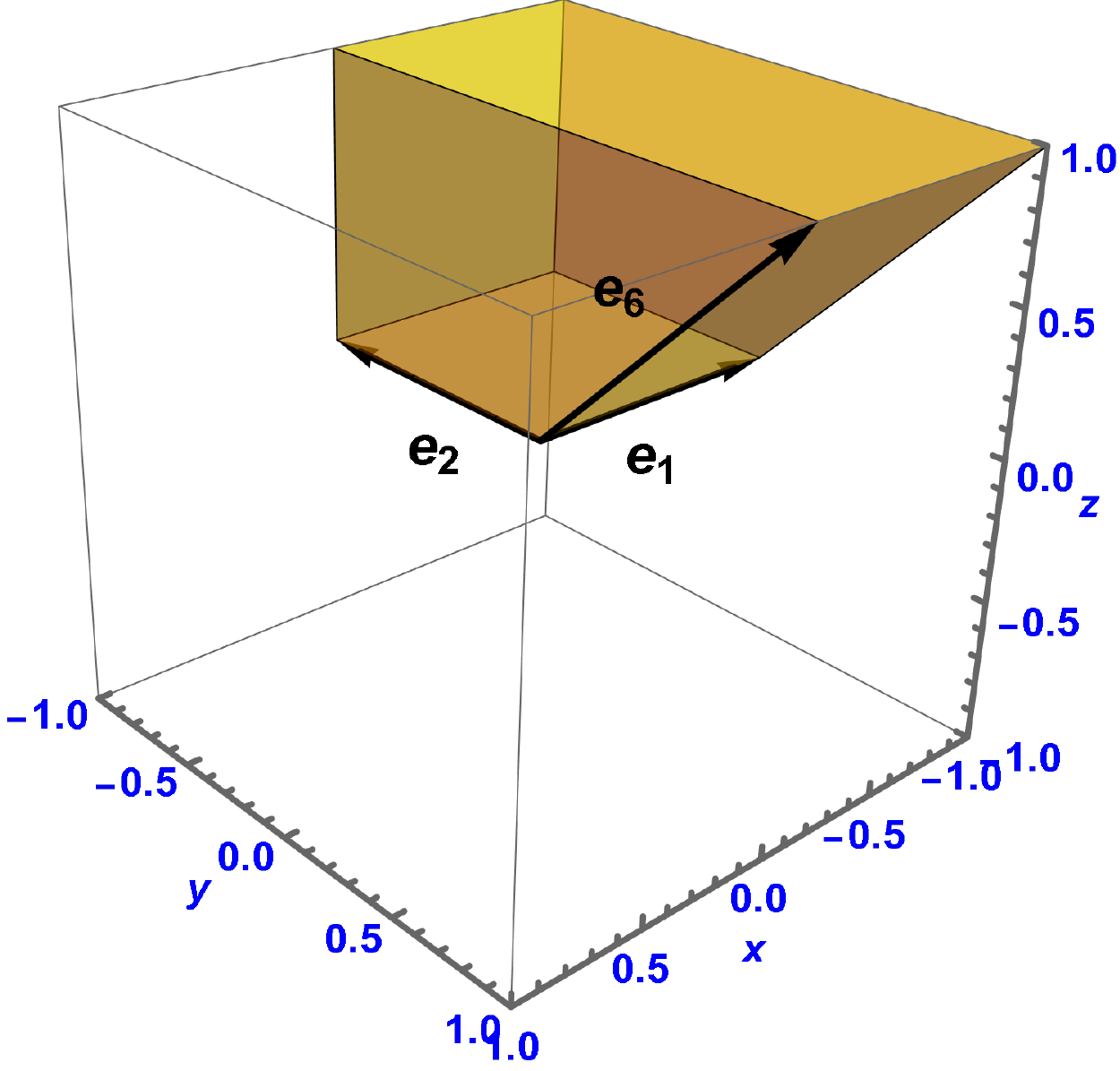}
\caption{\textit{Left:} Conic hull $C_{1,2,5}$ associated with the 3-combination $(1,2,5)$, generated by the vectors ${\bf e}_{1}=(-1,0,0), {\bf e}_{2}=(0,-1,0)$ and ${\bf e}_{5}=(1,0,1)$. \textit{Right:} Conic hull $C_{1,2,6}$ associated with the 3-combination $(1,2,6)$.}
\label{Hull_4_Boos}
\end{figure}

In this set of twenty $3$-combinations, we now have to keep only those whose associated conic hulls are 3-dimensional \cite{ABFGgeneral}. We recall here that the conic hull associated with the 3-combination  $(i_1,i_2,i_3)$ is generated by the three vectors ${\bf e}_{i_1}, {\bf e}_{i_2}$ and ${\bf e}_{i_3}$ (given in the third column of Table \ref{Table1}) and has its vertex at the origin. For example, the conic hull $C_{1,2,5}$ associated with the 3-combination $(1,2,5)$, coming from the gamma functions $\Gamma(-z_1),\Gamma(-z_2)$ and $\Gamma(a_2+z_1+z_3)$, is generated by the vectors $(-1,0,0)$, $(0,-1,0)$ and $(1,0,1)$. It is shown in Figure \ref{Hull_4_Boos}. 

It is easy to select the relevant 3-combinations by considering the $3 \times 3$ matrix $A_{i_1,i_2,i_3}$ associated with a given 3-combination $(i_1,i_2,i_3)$, defined as
\begin{equation} \label{Sign_Conformal}
     A_{i_1,i_2,i_3}=  \begin{pmatrix} {\bf e}_{i_1} \\ {\bf e}_{i_2} \\ {\bf e}_{i_3}\end{pmatrix}
\end{equation} 
It is indeed sufficient to omit those 3-combinations for which the matrix $A$ is singular because in that case the associated conic hulls cannot be 3-dimensional.  In the case of Eq.(\ref{MBtriangle}), we get 17 3-combinations, out of the 20 possible ones, with nonsingular matrix. The set $S$ of building blocks will thus also have a cardinal number equal to 17, as well as the set $S'$ of their associated conic hulls.

These sets read
\begin{align}
       S=\big\{ &B_{1,2,3}\,, B_{1,2,5} \,, B_{1,2,6}\,,B_{1,3,4}\,, B_{1,3,6} \,, B_{1,4,5}\,, B_{1,4,6}\,, B_{1,5,6}\,, B_{2,3,4}\,, B_{2,3,5}\,, B_{2,4,5}\,, B_{2,4,6}\,,\nonumber\\
        &B_{2,5,6}\,, B_{3,4,5}\,, B_{3,4,6}\,, B_{3,5,6}\,, B_{4,5,6}\big\}\label{Set_1_Building_Blocks}
   \end{align}
and
\begin{align}\label{CH_set1}
       S'=\big\{ &C_{1,2,3}\,, C_{1,2,5} \,, C_{1,2,6}\,,C_{1,3,4}\,, C_{1,3,6} \,, C_{1,4,5}\,, C_{1,4,6}\,, C_{1,5,6}\,, C_{2,3,4}\,, C_{2,3,5}\,, C_{2,4,5}\,, C_{2,4,6}\,,\nonumber\\
        &C_{2,5,6}\,, C_{3,4,5}\,, C_{3,4,6}\,, C_{3,5,6}\,, C_{4,5,6}\big\}
   \end{align}
We would like to open here a parenthesis concerning the cardinal number of $S$. 

It was advocated in the Yangian bootstrap analysis of conformal Feynman integrals in \cite{Loebbert:2019vcj} that there is a link between the MB building blocks and Yangian invariants. In the example of the massless one-loop conformal box integral and hexagon, one indeed obtains the same number of MB building blocks as the zeros of the fundamental solutions of the corresponding Yangian partial differential equations (see \cite{Loebbert:2019vcj} and \cite{Ananthanarayan:2020ncn}). 
However, we observed that there is no matching in several other examples, as for instance in the case of the massive one-loop conformal 3-point integral (and one-mass one-loop nonconformal 3-point integral) studied in \cite{Loebbert:2020hxk} where 29 (respectively 36) zeros are obtained from the Yangian bootstrap analysis whereas our MB analysis for these integrals gives only 17 (respectively 24) building blocks. What differentiate these two 3-point integrals from the box and hexagon is that the MB representations of the former own gamma functions in the denominator of the MB integrand whereas the MB representations of the latter do not.

We therefore believe that the discrepancy between the two methods is due to the fact that, in the Yangian approach, the authors of \cite{Loebbert:2020hxk} do not disentangle the gamma functions of the fundamental solution of the partial differential equations, that they all consider on an equal footing in their algorithm to find the zeros. 
However, our study of the same integrals from the MB side  clearly shows that, in fact, not all of these gamma functions should be retained: only those gamma functions that belong to the numerator of the corresponding MB integrand should be taken in the Yangian algorithm to obtain the zeros of the fundamental solution. Using this rule leads to a reduction of the number of relevant zeros, in the case mentioned above, from 29 to 17 and from 36 to 24, in agreement with our findings. This reduction of the number of zeros is also relevant in some other examples of \cite{Loebbert:2020glj}, as for instance the non-dual-conformal massive 2-point integral. 

Therefore, this suggests that the Yangian approach does not give the optimal number of zeros each time that the corresponding MB integral has gamma functions in the denominator of its integrand.

As an example, let us now explicitly show which zeros should not be taken into account in the Yangian analysis of the massive one-loop conformal 3-point integral. In this case, the fundamental solution of the recurrence relations obtained from the Yangian bootstrap approach is \cite{Loebbert:2020hxk}
\begin{align}
    f_{n_1,n_2,n_3}= &\frac{1}{\Gamma(1+n_1)\Gamma(1+n_2)\Gamma(1+n_3) \Gamma(1-a_1-n_1-n_2)\Gamma(1-a_2-n_1-n_3)} \nonumber \\ &\hspace{4.3cm} \times \frac{1}{\Gamma(1-a_3-n_2-n_3)\Gamma(\gamma+n_1+n_2+n_3)}
\end{align}
where $\gamma = D/2 +1/2$. 

As mentioned in \cite{Loebbert:2020hxk,Loebbert:2020glj} there are 29 triplets $(x,y,z)$ for which the following series is convergent:
\begin{align}
   \smashoperator[r]{ \sum_{(n_1,n_2,n_2) \in (x+\mathbb{Z},y+\mathbb{Z},z+\mathbb{Z})}} \hspace{0.5cm} f_{n_1,n_2,n_3} u^{n_1} v^{n_2} w^{n_3}  
\end{align}
The series associated with 17 of these triplets match with the 17 building blocks of Eq.\eqref{Set_1_Building_Blocks}. We have listed the remaining 12 triplets in Table \ref{Conformal_missing}.
\begin{table}[h!]
\centering
\begin{tabular}{||c c c||} 
 \hline
  & $(x,y,z)$ &    \\ [0.5ex] 
 \hline\hline
 $(0,0,1-\gamma)$ & $(0,1-\gamma,0)$ & $(1-\gamma,0,0)$ \\ 
 $(0,-a_1,1-\gamma+a_1)$ &  $(1-\gamma+a_3,-1+\gamma-a_1-a_3,1-\gamma+a_1)$ & $(0,1-\gamma+a_2,-a_2)$  \\
 $(1-\gamma+a_3,0,-a_3)$ & $(-1+\gamma-a_1-a_2,1-\gamma+a_2,1-\gamma+a_1)$ & $(-a_2,1-\gamma+a_2,0)$  \\
 $(1-\gamma+a_3,-a_3,0)$ & $(1-\gamma+a_3,1-\gamma+a_2,-1+\gamma-a_2-a_3)$ & $(-a_1,0,1-\gamma+a_1)$  \\ [1ex] 
 \hline
\end{tabular}
\caption{Triplets $(x,y,z)$ whose corresponding series is spurious.\label{Conformal_missing}}
\end{table}

A numerical analysis suggests that there is no contribution of the 12 series associated with the triplets of Table \ref{Conformal_missing}, in any of the series representations of the 3-point integral; therefore, we term them as spurious. We verified this as there is a perfect numerical match, at points which belong to the convergence regions of these spurious series, between the series representations built from the 17 building blocks obtained from our approach, and the Feynman parametrization. 

We also performed an analytic check, by showing that the 3 spurious series associated with $(0,1-\gamma+a_2,-a_2)$, $(1-\gamma+a_3,0,-a_3)$ and $(1-\gamma+a_3,1-\gamma+a_2,-1+\gamma-a_2-a_3)$ converge for values of $u,v,w$ which belong to the region of convergence of the RHS of Eq.(\ref{Srivastava_continuation}). This cannot be the case except if their overall coefficients are null. Indeed, Eq.(\ref{Srivastava_continuation}) is a well-known result (see Eq.(62) p.293 in \cite{Srivastava}) built from two of the 17 building blocks (as explained below).

For the non-dual-conformal massive 2-point integral analyzed in \cite{Loebbert:2020glj}, we obtain 8 building blocks, to be compared to the 13 zeros mentioned in \cite{Loebbert:2020glj}. 
In the same way as above, it is easy to show that 5 of these 13 zeros are spurious. One of the spurious zeros is associated with the doublet $(0,-\alpha-\beta)$, given in Eq.(7.19) of \cite{Loebbert:2020glj} where a numerical analysis has explicitly shown that it was indeed not contributing.

From the above we conclude that our MB method gives a way to directly identify, in general, those of the Yangian zeros that are relevant and to remove the spurious ones.

Let us close here this parenthesis and come back to the explicit computation of the series representations associated with the MB integral in Eq.(\ref{MBtriangle}).
\\

The next step to obtain these series representations is to find the largest subsets of conic hulls in $S'$ whose intersection is nonempty, as we have observed that there is a one-to-one correspondence between these subsets and the series representations \cite{ABFGgeneral}. A straightforward geometrical analysis yields 14 such subsets, which therefore leads to 14 series representations that are analytic continuations of one another. 

The 14 subsets are
\begin{align}
&\{C_{1,2,3}\}\nonumber\\\nonumber\\
& \{C_{1,2,5},C_{1,2,6}\}, \{C_{1,3,4},C_{1,3,6}\}, \{C_{2,3,4},C_{2,3,5}\}\nonumber\\\nonumber\\
&\{C_{1,2,5},C_{2,4,6},C_{2,5,6}\}, 
\{C_{1,2,6},C_{1,4,5},C_{1,5,6}\}, 
\{C_{1,3,4},C_{3,4,6},C_{3,5,6}\}, 
\{C_{1,3,6},C_{1,4,5},C_{1,4,6}\}\nonumber\\ 
&\{C_{2,3,4},C_{3,4,5},C_{3,5,6}\}, 
\{C_{2,3,5},C_{2,4,5},C_{2,4,6}\},\nonumber\\\nonumber\\
&\{C_{1,4,5},C_{1,4,6},C_{3,5,6},C_{3,4,6}\},
\{C_{1,4,5},C_{1,5,6},C_{2,4,6},C_{2,5,6}\}, \{C_{1,4,5},C_{2,4,6},C_{3,5,6},C_{4,5,6}\} \nonumber\\\nonumber\\
&  \{C_{2,4,5},C_{2,4,6},C_{3,4,5},C_{3,5,6}\}\label{List_inter}
\end{align}

In the list above we have classified the subsets of conic hulls in 5 different rows. The second, third and fourth rows contain subsets that are linked together by symmetry. Indeed, due to the symmetry of the MB representation in Eq.\eqref{MBtriangle} which reflects the symmetry of the $H_C$ function, 9 of the 14 corresponding series representations can be obtained from the series representations $S_2, S_3$ and $S_4$ which belong to the following set of 5 independent series representations 
   \begin{equation}\label{3point_Series_Representation}
 J(a_1,a_2,a_3;u,v,w)=\left\{\begin{array}{ll}
S_1=B_{1,2,3} &   \hspace{0.8cm} (\text{Region $\mathcal{R}_1$})\\
S_2=B_{1,2,5}+B_{1,2,6} &  \hspace{0.8cm} (\text{Region $\mathcal{R}_2$})\\
S_3=B_{1,2,5}+B_{2,4,6}+B_{2,5,6} &  \hspace{0.8cm} (\text{Region $\mathcal{R}_3$}) \\
S_4=B_{1,4,5}+B_{1,4,6}+B_{3,4,6}+B_{3,5,6}&  \hspace{0.8cm} (\text{Region $\mathcal{R}_4$})\\
S_5=B_{1,4,5}+B_{2,4,6}+B_{3,5,6}+B_{4,5,6} &  \hspace{0.8cm} (\text{Region $\mathcal{R}_5$})
\end{array}\right.
\end{equation}

The last step to obtain the expressions of the series representations is to compute, from the set of poles associated with each 3-combination corresponding to the building blocks, the related residues.

The set of poles associated with the 3-combination $(i_1,i_2,i_3)$ is the solution, in terms of $z_i$, of the linear system 
 \begin{equation}\label{Poles_Associated}
     \begin{pmatrix} s_{i_1} \\ s_{i_2} \\ s_{i_3}\end{pmatrix}=  \begin{pmatrix} -n_1 \\ -n_2 \\ -n_3\end{pmatrix}, \hspace{0.5cm} n_i\in \mathbb{N},  (i=1,2,3)
 \end{equation} 
where $s_{i_1}$ is the argument of the gamma function in the numerator with label $i_1$.

As an example, the set of poles corresponding to $(1,2,5)$ is at $(z_1,z_2,z_3)=(n_1,n_2,-a_2-n_1-n_3)$.

The steps to proceed are now to perform changes of variables on the MB integrand so as to bring the singularities to the origin, and apply the generalized reflection formula 
\begin{align}
\Gamma(z-n)=\frac{\Gamma(1+z)\Gamma(1-z)(-1)^n}{z\,\,\Gamma(n+1-z)}, \hspace{1.5cm} n\in \mathbb{Z}
\end{align} 
to explicitly extract the singular factors of the gamma functions that are singular at the origin. It is then enough to divide the obtained expression by $\vert\text{det}A\vert$ and put $z_1=z_2=z_3=0$ after removing the singular factors from the denominator. Summing over the $n_i$, $(i=1,2,3)$ and repeating the procedure for each relevant 3-combination, one then gets the desired expressions of all the building blocks. These expressions are given in Appendix \ref{BBset1}.

Taking into account the overall factor of Eq.(\ref{overall}), $S_1=B_{1,2,3}$ reproduces the $H_C$ result of \cite{Loebbert:2020hxk}, which reads
\begin{align}
 I_{3 \bullet}^{m_{1} m_2 m_{3}}=\frac{\pi^{D/2+1/2}}{2^{D-1}m_1^{a_1}m_2^{a_2}m_3^{a_3}\Gamma(D/2+1/2)}\pFq{}{H}{C}{a_1,a_2,a_3}{\frac{D+1}{2}}{u,v,w}
\end{align}

As another example, the substitution $(a_1,a_2,a_3,D/2+1/2) \to(\alpha,\beta,\beta',\gamma)$ into $S_2$ gives the analytic continuation formula 
\begin{align}\label{Srivastava_continuation}
  \pFq{}{H}{C}{\alpha,\beta,\beta'}{\gamma}{u,v,w} &= \frac{\Gamma(\gamma)\Gamma(\beta'-\beta)}{\Gamma(\gamma-\beta)\Gamma(\beta')}(-w)^{-\beta}\pFq{}{G}{C}{\beta,\alpha,\beta-\gamma+1}{\beta-\beta'+1}{\frac{u}{w},\frac{1}{w},v}\nonumber \\ & + \frac{\Gamma(\gamma)\Gamma(\beta-\beta')}{\Gamma(\gamma-\beta')\Gamma(\beta)}(-w)^{-\beta'}\pFq{}{G}{C}{\beta',\alpha,\beta'-\gamma+1}{\beta'-\beta+1}{\frac{v}{w},\frac{1}{w},u}
\end{align}
where
\begin{align}\label{G_C definition}
  \pFq{}{G}{C}{\alpha,\beta,\beta'}{\gamma}{u,v,w} =\sum\limits_{n_1,n_2,n_3=0}^{\infty}\frac{(\alpha)_{n_1+n_2}(\beta)_{n_1+n_3}(\beta')_{n_2-n_3}}{(\gamma)_{n_1+n_2-n_3}} \frac{u^{n_1}v^{n_2}w^{n_3}}{n_1! \, n_2! \, n_3!}
\end{align}
As already said, Eq.(\ref{Srivastava_continuation}) is known (see Eq.(62) p.293 in \cite{Srivastava}). It thus provides an analytic check of our method. 

To our knowledge the other series representations $S_3, S_4, S_5,...$ provide new analytic continuations of the $H_C$ series. All have been checked by a comparison with the direct numerical evaluation of the Feynman parameterization of the 3-point integral. Each time an excellent agreement has been obtained.

Let us now turn to the derivation of the convergence regions $\mathcal{R}_i (i=1,\cdots,5)$. As explained in \cite{ABFGgeneral}, it can be very helpful to rely on the concept of master series in order to simplify the convergence analysis. Indeed, each of the series representations in Eq.(\ref{3point_Series_Representation}) has an associated master series which, in the case of the massive one-loop conformal 3-point integral, directly gives the convergence of the corresponding series representation, which avoids the convergence study of all the building blocks that make a given series representation. We proved this fact for $S_i, (i=1,\cdots,4)$ but not for $S_5$ whose convergence region $\mathcal{R}_5$ is harder to compute: in the latter case we only give a conjecture of the expression of $\mathcal{R}_5$ as well as a proof of the maximal volume that it can fill (see Appendix \ref{conv_set1}). 

\begin{figure}[h]
\centering
\includegraphics[width=6.6cm]{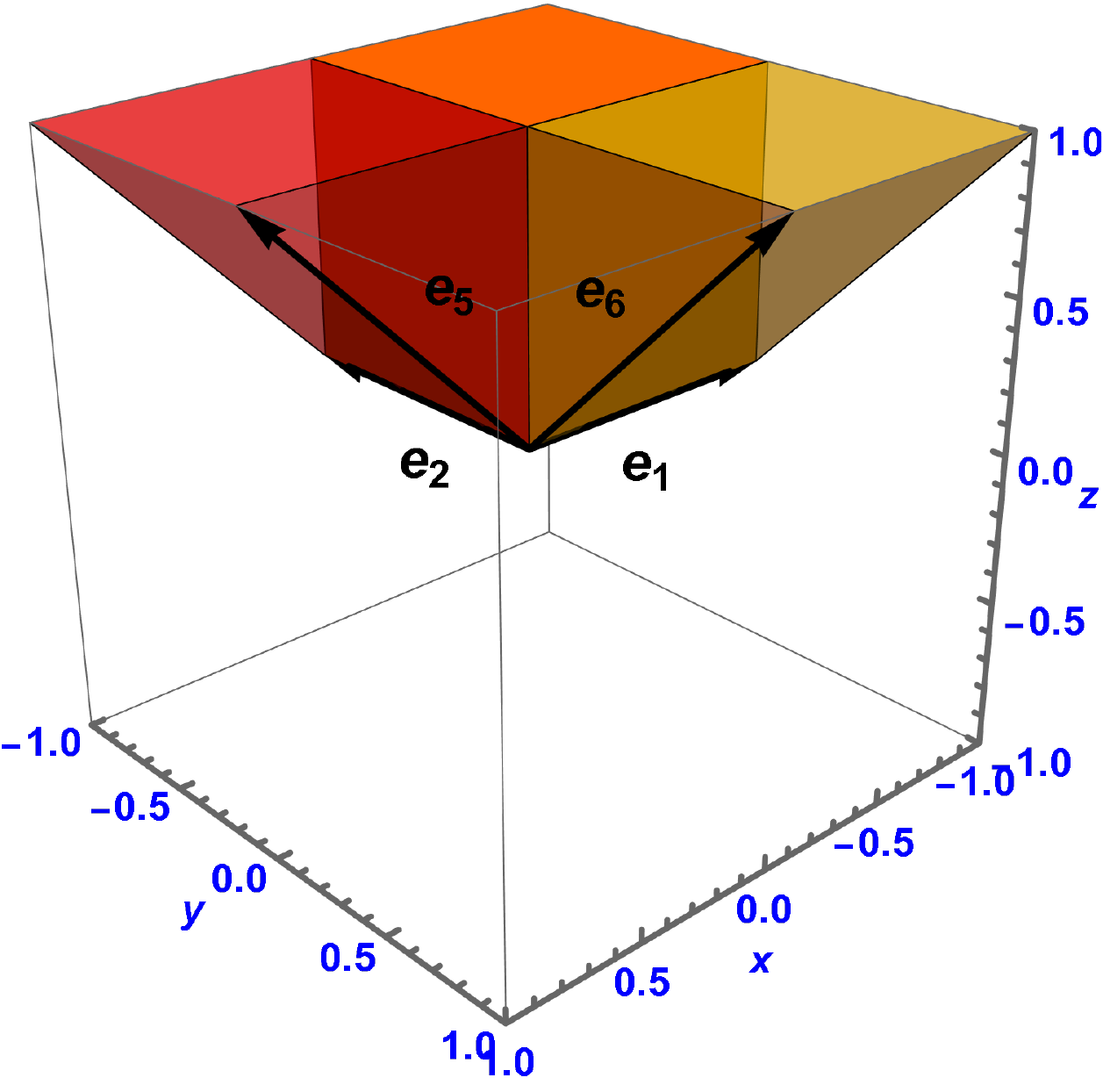}
\caption{Intersection of the conic hulls shown in Fig.\ref{Hull_4_Boos}. The intersection region is shown in orange and corresponds to the master conic hull of $S_2$ in Eq.(\ref{3point_Series_Representation}).\label{3MB_Boos_Cone2}}
\end{figure}

It can happen that the master series corresponds to one of the series involved in the series representation. For example, the master series of $S_5$ coincides with $B_{4,5,6}$. In contrast to $S_5$, neither $B_{1,2,5}$ nor $B_{1,2,6}$ is the master series of $S_2$. To derive the master series of $S_2$ one has to find the generating vectors of the corresponding master conic hull which, as shown in orange in Fig.\ref{3MB_Boos_Cone2}, is the intersection of the conic hulls $C_{1,2,5}$ and $C_{1,2,6}$.
These vectors are $(-c_1,0,0)$,$(0,-c_2,0)$ and $(0,0,c_3)$, where $c_1,c_2,c_3>0$. From these vectors one then builds a 3-combination of gamma functions, here $(\Gamma(-c_1z_1),\Gamma(-c_2z_2),\Gamma(c_3z_3))$. The constraints on the values of the $c_i$ come from the fact that the corresponding set of poles, $(n_1/c_1,n_2/c_2,-n_3/c_3)$,  has to be able to describe both the sets of poles associated with $(1,2,5)$ and $(1,2,6)$, after suitable changes of variables. Here this leads to $c_1=c_2=c_3=1$. Having found that the set of poles associated with the master series is $(n_1,n_2,-n_3)$ it is now straightforward to obtain the form of the master series using the same procedure as described above for the last step of the derivation of the series representations.

Once the master series is obtained, one can compute its convergence region using Horn's theorem \cite{Srivastava}. It is possible to check that the obtained result is correct by a direct study of the convergence region of each of the building blocks involved in the series representations, still from Horn's theorem.

\subsection{Set 2: Series Representations\label{Series_set2}}

A comparison of the MB representations in Eqs.(\ref{MBtriangle}) and (\ref{Conformal_Triangle_2}) shows that our method to find the series representations of Eq.(\ref{Conformal_Triangle_2}) will give the same set of conic hulls $S'$ as the one shown in Eq.(\ref{CH_set1}). Therefore, as in the previous case, one can find 14 different series representations of Eq.(\ref{Conformal_Triangle_2}). These are linked together by the same symmetry properties as the 14 series representations of Eq.(\ref{MBtriangle}). 

Therefore, only 5 are independent and read
   \begin{equation}\label{3point_Series_Representation_2}
 J_2(a_1,a_2,a_3;u,v,w)=\left\{\begin{array}{ll}
S'_1=B'_{1,2,3} &   \hspace{0.8cm} (\text{Region $\mathcal{R}'_1$})\\
S'_2=B'_{1,2,5}+B'_{1,2,6} &  \hspace{0.8cm} (\text{Region $\mathcal{R}_2$})\\
S'_3=B'_{1,2,5}+B'_{2,4,6}+B'_{2,5,6} &  \hspace{0.8cm} (\text{Region $\mathcal{R}'_3$}) \\
S'_4=B'_{1,4,5}+B'_{1,4,6}+B'_{3,4,6}+B'_{3,5,6}&  \hspace{0.8cm} (\text{Region $\mathcal{R}'_4$})\\
S'_5=B'_{1,4,5}+B'_{2,4,6}+B'_{3,5,6}+B'_{4,5,6} &  \hspace{0.8cm} (\text{Region $\mathcal{R}'_5$})
\end{array}\right.
\end{equation}
where the $B'_{i,j,k}$ and the $\mathcal{R}'_i$ are given in Appendix \ref{BBset2} and \ref{conv_set2}, respectively.

\section{A set of new quadratic transformations for a class of multiple hypergeometric series\label{quadratic}}

In our first attempt to derive the convergence region of $S'_1$ in Eq.(\ref{3point_Series_Representation_2}), we have split each of the three sums into sums over odd and even terms, in order to simplify the convergence analysis. This indeed allowed us to get rid of the 1/2 factors in the arguments of the gamma functions, thereby obtaining an alternative expression which only involves Srivastava's $H_B$ triple hypergeometric series \cite{Srivastava64,Srivastava} 
\begin{align}
   &\sum_{k,l,n=0}^\infty(\hat a_1)_{\hat k+\hat l}(\hat a_2)_{\hat k+\hat n}(\hat a_3)_{\hat l+\hat n}\frac{(-2u')^k(-2v')^l(-2w')^n}{k!l!n!}=\nonumber
  \\& \frac{\pi^{3/2}}{\Gamma(\hat a_1)\Gamma(\hat a_2)\Gamma(\hat a_3)}\left\{\frac{\Gamma(\hat a_1)\Gamma(\hat a_2)\Gamma(\hat a_3)}{\Gamma(\frac{1}{2})^3} \pFq{}{H}{B}{\hat a_1,\hat a_2,\hat a_3}{\frac{1}{2},\frac{1}{2},\frac{1}{2}}{(2u-1)^2,(2v-1)^2,(2w-1)^2}\right.\nonumber\\&
 + \left.\left(2u-1\right)\frac{\Gamma(\hat a_1+\frac{1}{2})\Gamma(\hat a_2+\frac{1}{2})\Gamma(\hat a_3)}{\Gamma(\frac{1}{2})^2\Gamma(\frac{3}{2})} \pFq{}{H}{B}{\hat a_1+\frac{1}{2},\hat a_2+\frac{1}{2},\hat a_3}{\frac{3}{2},\frac{1}{2},\frac{1}{2}}{(2u-1)^2,(2v-1)^2,(2w-1)^2}\right.\nonumber
 \\& +\text{Perm.}\nonumber
 \\&+(2u-1)(2v-1)\frac{\Gamma(\hat a_1+1)\Gamma(\hat a_2+\frac{1}{2})\Gamma(\hat a_3+\frac{1}{2})}{\Gamma(\frac{1}{2})\Gamma(\frac{3}{2})^2} \nonumber
 \\&\hspace{2cm}\times\pFq{}{H}{B}{\hat a_1+1,\hat a_2+\frac{1}{2},\hat a_3+\frac{1}{2}}{\frac{3}{2},\frac{3}{2},\frac{1}{2}}{(2u-1)^2,(2v-1)^2,(2w-1)^2}\nonumber
 \\& +\text{Perm.}\nonumber
 \\& +(2u-1)(2v-1)(2w-1)\frac{\Gamma(\hat a_1+1)\Gamma(\hat a_2+1)\Gamma(\hat a_3+1)}{\Gamma(\frac{1}{2})^3} \nonumber
 \\&\hspace{2cm}\left.\times\pFq{}{H}{B}{\hat a_1+1,\hat a_2+1,\hat a_3+1}{\frac{3}{2},\frac{3}{2},\frac{3}{2}}{(2u-1)^2,(2v-1)^2,(2w-1)^2}\right\}\label{HBexp}
\end{align}
where in the RHS we have replaced $u'$ by $(1-2u)$, etc. and where $\hat a_i\doteq a_i/2, (i=1,2,3)$.

The convergence properties of $H_B$ are well known \cite{Srivastava}. It is therefore straightforward to conclude that the convergence region of the LHS of Eq.(\ref{HBexp}) is
\begin{equation}
\mathcal{R}'_1=\vert 2u-1\vert^2+\vert 2v-1\vert^2+\vert 2w-1\vert^2+2\vert 2u-1\vert\vert 2v-1\vert\vert 2w-1\vert<1
\end{equation}
Now, since $\mathcal{R}'_1\subset \mathcal{R}_1$ where $\mathcal{R}_1$, defined in Eq.(\ref{R1set1}), is the convergence region of the $H_C$ triple series, one can obtain a nice (and we believe new) quadratic transformation of $H_C$ in terms of $H_B$ by using the fact that, in $\mathcal{R}'_1$, $S_1=\frac{2^{D-1}}{4\sqrt{\pi}}S'_1$ which allows us to write
\begin{align}
\pFq{}{H}{C}{a_1,a_2,a_3}{\frac{D}{2}+\frac{1}{2}}{u,v,w}=&\frac{\pi^{3/2}}{\Gamma(\frac{D}{2})}\frac{\Gamma(D)2^{1-D}}{\Gamma(\hat a_1+\frac{1}{2})\Gamma(\hat a_2+\frac{1}{2})\Gamma(\hat a_3+\frac{1}{2})}\nonumber\\
 &\times\sum_{k,l,n=0}^\infty(\hat a_1)_{\hat k+\hat l}(\hat a_2)_{\hat k+\hat n}(\hat a_3)_{\hat l+\hat n}\frac{(-2u')^k(-2v')^l(-2w')^n}{k!l!n!}\label{linkHC}
\end{align}
The quadratic transformation of $H_C$ is then obtained by replacing the series in the RHS of Eq.(\ref{linkHC}) by its transformed expression given in Eq.(\ref{HBexp}) and it is valid only in $\mathcal{R}'_1$.

This quadratic transformation is the extension of a lowest-order quadratic transformation, which can be deduced from a confrontation of the massive one-loop conformal 2-point integral results given in Eq.(8.23) and Eqs.(8.9) and (8.18) of \cite{Loebbert:2020glj}. Indeed, after using the splitting of odd and even terms on Eq.(8.18) one obtains 
\begin{align}\label{Erdelyi}
 &\pFq{2}{F}{1}{a_1,a_2}{\frac{D+1}{2}}{u}
=\frac{\sqrt{\pi}\Gamma(D)}{2\Gamma(\frac{D}{2})\Gamma(a_1)\Gamma(a_2)}\left\{\frac{\Gamma(\hat a_1)\Gamma(\hat a_2)}{\Gamma(\frac{1}{2})}\pFq{2}{F}{1}{\hat a_1,\hat a_2}{\frac{1}{2}}{(2u-1)^2} \right.\nonumber\\
&\left.\hspace{2.85cm}+(2u-1)\frac{\Gamma\left(\hat a_1+\frac{1}{2}\right)\Gamma\left(\hat a_2+\frac{1}{2}\right)}{\Gamma(\frac{3}{2})} \pFq{2}{F}{1}{\hat a_1+\frac{1}{2},\hat a_2+\frac{1}{2}}{\frac{3}{2}}{(2u-1)^2}\right\}
\end{align}
where $D=a_1+a_2$.

Eq.(\ref{Erdelyi}) is nothing but a well-known quadratic transformation of the Gauss $_2F_1$ hypergeometric series (see Eq.(28) p.65 of \cite{Erdelyi}). As for the case of $H_C$ above, the convergence region of the RHS of Eq.(\ref{Erdelyi}) is included in the convergence region of the LHS, thereby giving validity for the quadratic transformation only in the former.

Having obtained the $n=2$ and $n=3$ quadratic transformations for $_2F_1$ and $H_C$, it is natural to expect that the expressions of the massive one-loop conformal $n$-point integral given in Eqs.(\ref{conjecture_regA}) and (\ref{conjecture_regB}) are related by a general quadratic transformation which includes both cases above as particular cases and which reads
\begin{align}\label{conj_quad}
   \sum_{N_{12},N_{13},\cdots,N_{n-1,n}=0}^\infty  &\frac{\prod_{i=1}^{n}\left(a_i\right)_{\sum_{\alpha \in B_{n|i}}N_\alpha}}{\left( \frac{D+1}{2}\right)_{\sum_{\alpha \in B_{n}}N_\alpha}}\prod_{\alpha\in B_{n} } \left(\frac{u_{\alpha}^{N_\alpha}}{N_\alpha!}\right)= \hspace{0.5cm}C \hspace{0.2cm}\smashoperator{ \sum_{s_{12},s_{13},\ldots,s_{n-1,n}=0}^{1}} \hspace{1.3cm} \smashoperator{\prod_{\alpha \in B_n}}\left((-v_{\alpha})^{s_{\alpha}}\right) \nonumber\\
    &\times H_S^{(n)}\pFq{}{}{}{\hat{a}_1+\smashoperator{\sum_{\alpha\in B_{n|1}}} \hat{s}_{\alpha},\cdots,\hat{a}_n+\smashoperator{\sum_{\alpha\in B_{n|n}}} \hat{s}_{\alpha}\,}{\frac{1}{2}+s_{12},\frac{1}{2}+s_{13},\cdots,\frac{1}{2}+s_{n-1,n}}{ v_{12}^{2},v_{13}^{2},\cdots,v_{n-1,n}^{2}}
\end{align}
where $v_{\alpha}=1-2u_{\alpha}$, $D=\sum_{i=1}^na_i$, 
\begin{align}
   &H_S^{(n)}\pFq{}{}{}{a_1,\cdots,a_n }{b_{12},b_{13},\cdots,b_{n-1,n}}{ x_{12},x_{13},\cdots,x_{n-1,n}}=\nonumber\\
   &\hspace{6cm} \smashoperator{\sum_{k_{12},k_{13},\ldots,k_{n-1,n}=0}^{\infty}} \hspace{0.5cm} \frac{\prod_{j=1}^{n}\Gamma(a_{j}+{\sum_{\alpha\in B_{n|j}} k_{\alpha}})}{\prod_{\alpha\in B_n}\Gamma(b_{\alpha}+{k_{\alpha}})}\prod_{\alpha\in B_n}\frac{(x_{\alpha})^{k_{\alpha}}}{k_{\alpha}!} 
\end{align}
and
\begin{align}
   C= \frac{\pi^{\frac{n(n-1)}{4}}\Gamma(D)}{2^{n-1}\Gamma(\frac{D}{2})\prod_{j=1}^{n}\Gamma(a_j)}
\end{align}
A study of the convergence regions of both the LHS and RHS of Eq.(\ref{conj_quad}) is needed to complete the proof of this tower of quadratic transformations.

Note that this set of new quadratic transformations, which at lowest order gives the known result Eq.(\ref{Erdelyi}) involving the $_2F_1$ hypergeometric function, is a consequence of choosing different conformal variables in the Feynman parameterization of the massive one-loop conformal $n$-point function.

We conclude from this analysis that QFT can be used as a tool to derive nontrivial results in hypergeometric functions theory.

\section{Resonant case: unit propagator powers \label{resonant}}
We now redo the exercise of Section \ref{Set1series} for unit propagator powers, i.e. $a_i=1$ for $i=1,2,3$. This case is harder to treat except for the simplest series representation which in fact is nonresonant and for which it is straightforward to obtain a result already derived in Eqs.(9.14) and (9.16) of \cite{Loebbert:2020glj}. We denote this result as $S_1^{\text{unit}}$ in the following.

In passing, we conjecture the following amusing relation \begin{equation}
\pFq{}{H}{C}{1,1,1}{2}{\frac{1}{2},\frac{1}{2},\frac{1}{2}}=\pi
\end{equation}

Let us now turn to a less simple case, by focusing on what becomes the equivalent, for the unit propagator powers case, of $S_2$ in Eq.(\ref{3point_Series_Representation}) and which we denote as $S_2^{\text{unit}}$.
As seen in Section \ref{Set1series} the two sets of poles that are associated to this series representation are those of the 3-combinations $(1,2,5)$ and $(1,2,6)$, which are located at $(n_1,n_2,-1-n_1-n_3)$ and $(n_1,n_2,-1-n_2-n_3)$. In contrast with the generic propagator powers case, some poles in these two sets are now overlapping, which is the typical situation met in the resonant case where poles of order greater than 1 are expected. One thus cannot use the simple nonresonant procedure based on building blocks, which has been used in the previous sections, for these poles of higher multiplicity. One instead has to follow the more general procedure described in \cite{ABFGgeneral} for the resonant case. In the calculation process, one must also be careful to consider overlapping poles only once. It is possible to detect such poles easily by noting that for overlapping poles the number of singular gamma functions in the numerator is greater than three.

Let us now dive into the details of the calculations, and consider each of the sets of poles mentioned above one at a time.\\

     {\bf Set of poles I:} $(n_1,n_2,-1-n_1-n_3)$\\
     
     These are the poles coming from the 3-combination $(1,2,5)$.
    
    Let us shift these poles to the origin by the change of variable $z_i \to z_i+n_i$ for $i=1,2$ and $z_3\to z_3-1-n_1-n_3$. The integrand of \eqref{MBtriangle} becomes
    \begin{align}\label{Cone_21_Singular}
        & \left(-u\right)^{z_{1}+n_1} \left(-v\right)^{z_{2}+n_2}  \left(-w\right)^{z_{3}-1-n_1-n_3}
    \frac{\Gamma (-z_{1}-n_1)\Gamma(-z_{2}-n_2)\Gamma(-z_{3}+1+n_1+n_3)}{\Gamma(1+z_1+z_2+z_3+n_2-n_3)}
   \nonumber \\ & \times {\Gamma}(1+z_{1}+z_{2}+n_1+n_2){\Gamma}(z_{1}+z_{3}-n_3){\Gamma}(z_{2}+z_{3}+n_2-n_1-n_3)
    \end{align}
One observes that the poles at the origin are not of the same type for all values of $n_i$. Therefore, one has to treat each type of poles separately.\\

   $\bullet$ Type 1: $n_2\geq 1+n_1+n_3$ and remaining summation variables run from $0$ to $\infty$.\\\\ For this type of poles only the first, second and fifth gamma functions of the numerator of (\ref{Cone_21_Singular}) are singular at the origin; therefore, one can proceed as in the nonresonant case to obtain the series representation. Dividing by $\vert \text{det} A_{1,2,5}\vert$,
applying the generalized reflection formula on each of the singular gamma functions and removing the singular factors, one gets the analytic part of the integrand
    \begin{align}
       &(-1)^{n_1+n_2+n_3}\left(-u\right)^{z_{1}+n_1} \left(-v\right)^{z_{2}+n_2}  \left(-w\right)^{z_{3}-1-n_1-n_3}
    \frac{\Gamma (1-z_{1})\Gamma (1+z_{1})\Gamma(1-z_{2})\Gamma(1+z_{2})}{\Gamma(1+z_1+n_1)\Gamma(1+z_2+n_2)}
   \nonumber \\ & \times \frac{\Gamma(-z_{3}+1+n_1+n_3)\Gamma(1+z_{1}+z_{2}+n_1+n_2)\Gamma(1+z_{1}+z_{3})\Gamma(1-z_{1}-z_{3})}{\Gamma(1-z_{1}-z_{3}+n_3)\Gamma(1+z_1+z_2+z_3+n_2-n_3)} \nonumber \\ & \times \Gamma(z_{2}+z_{3}+n_2-n_1-n_3)
    \end{align} 
which by putting $z_1=z_2=z_3=0$ and summing over the $n_i$ gives the contribution
\begin{align}\label{Arbitrary-Cone-2-1-Series}
    S^{\text{unit}}_{2,1}= 
   \frac{v}{w} \sum_{n_1,n_2,n_3=0}^{\infty}(-1)^{1+n_1}\frac{{\Gamma}^{2}(1+n_2)\Gamma(1+n_1+n_3)\Gamma(2+2n_1+n_2+n_3)}{\Gamma(2+n_1+n_2)\Gamma(2+n_1+n_2+n_3)} \frac{(uv/w)^{n_1}v^{n_2}(v/w)^{n_3}}{n_1!\,n_2!\, n_3! } 
\end{align}
where we shifted the summation variables to keep the limit from $0$ to $\infty$ in each $n_i$.\\

$\bullet$ Type 2: $n_3\leq n_2\leq n_1+n_3$ and remaining summation variables run from $0$ to $\infty$.\\\\ For this type of poles the first, second and fifth but also sixth gamma functions of the numerator of (\ref{Cone_21_Singular}) are singular at the origin. As there are more than three singular gamma functions in the numerator, this indicates that these poles overlap with another set of poles associated with the series representation that we look for. This is a resonant case, for which the multivariate residues approach cannot be avoided. In particular, the grouping of the singular factors, needed for the transformation law, will have to be performed \cite{ABFGgeneral}.

As $\Gamma(a_3+z_2+z_3)$ is singular, this suggests that the considered type of poles overlaps with the poles from the 3-combination $(1,2,6)$ which, as said above, is also associated with $S_2^{\text{unit}}$. Obviously here there is no other choice because we know that $S_2^{\text{unit}}$ is built from only $(1,2,5)$ and $(1,2,6)$. At this level, one has to remember that, when one will evaluate the contributions coming from $(1,2,6)$ later, the overlapping poles that we consider presently will have to be omitted, in order to avoid double counting. 

Applying the generalized reflection formula on each singular gamma functions in \eqref{Cone_21_Singular} one obtains, for the MB integrand, the expression
    \begin{align}
       & \frac{\left(-u\right)^{z_{1}+n_1} \left(-v\right)^{z_{2}+n_2}}{(-z_1)(-z_2)(z_1+z_3)(z_2+z_3)}  \left(-w\right)^{z_{3}-1-n_1-n_3}
    \frac{\Gamma (1-z_{1})\Gamma (1+z_{1})\Gamma(1-z_{2})\Gamma(1+z_{2})}{\Gamma(1+z_1+n_1)\Gamma(1+z_2+n_2)}
   \nonumber \\ & \times \frac{\Gamma(-z_{3}+1+n_1+n_3)\Gamma(1+z_{1}+z_{2}+n_1+n_2)\Gamma(1+z_{1}+z_{3})\Gamma(1-z_{1}-z_{3})}{\Gamma(1-z_{1}-z_{3}+n_3)\Gamma(1+z_1+z_2+z_3+n_2-n_3)} \nonumber \\ & \times \frac{\Gamma(1+z_{2}+z_{3})\Gamma(1-z_{2}-z_{3})}{\Gamma(1-z_{2}-z_{3}+n_1+n_3-n_2)}
    \end{align}
which explicitly shows the singular factors in the denominator.

One then finds the grouping of singular factors following the algorithm of \cite{ABFGgeneral}, by writing
\begin{align}
    \left( f_1 ,f_2, f_3 \right)&=S_{1,2,5}\left(-z_1,-z_2,z_1+z_3 \right)+S_{1,2,6}\left( -z_1,-z_2,z_2+z_3 \right)\nonumber\\&=\left( -z_1,-z_2,z_1+z_3 \right)+\left( -z_1,-z_2,z_2+z_3 \right)=\left(-z_1,-z_2,(z_1+z_3)(z_2+z_3)\right)
\end{align}
Thus one has the following grouping $\left( f_1 ,f_2, f_3 \right)=\left( -z_1,-z_2,(z_1+z_3)(z_2+z_3) \right)$ on which the transformation law can now be applied. This can be done automatically using the \textit{Mathematica} package $\texttt{MultivariateResidues}$ \cite{Larsen:2017aqb}, from which we get the contribution
\begin{align}
    S^{\text{unit}}_{2,2}= & 
    \frac{1}{w}\sum_{n_1,n_2,n_3=0}^{\infty}(-1)^{1+n_2}\frac{{\Gamma}(1+n_1+n_2+n_3)\Gamma(1+n_1+2n_2+n_3)}{\Gamma(1+n_1+n_2)\Gamma(1+n_2+n_3)} \bigg( \text{log} (-w) +\psi(1+n_1)  \nonumber \\ & -\psi(1+n_2)+\psi(1+n_3)-\psi(1+n_1+n_2+n_3) \bigg) \frac{(u/w)^{n_1}(uv/w)^{n_2}(v/w)^{n_3}}{n_1!\,n_2!\, n_3! } 
\end{align}
where we shifted the summation variables to keep the limit from $0$ to $\infty$ in each $n_i$.

$\bullet$ Type 3: $n_2\leq n_3-1$, $n_3$ runs from $1$ to $\infty$ and $n_1$ from $0$ to $\infty$.\\\\ We shift $n_3 \to n_3+1$ for convenience. This type of poles also have the first, second, fifth and sixth gamma functions singular at the origin. But it is different from Type 2 as now the denominator also is singular at the origin.  Applying the generalized reflection formula on each singular gamma function in \eqref{Cone_21_Singular} we get the analytic part
    \begin{align}
       & (-1)^{n_2+n_3}\left(-u\right)^{z_{1}+n_1} \left(-v\right)^{z_{2}+n_2}  \left(-w\right)^{z_{3}-2-n_1-n_3}
    \frac{\Gamma (1-z_{1})\Gamma (1+z_{1})\Gamma(1-z_{2})\Gamma(1+z_{2})}{\Gamma(1+z_1+n_1)\Gamma(1+z_2+n_2)}
   \nonumber \\ & \times \frac{\Gamma(-z_{3}+2+n_1+n_3)\Gamma(1+z_{1}+z_{2}+n_1+n_2)\Gamma(1+z_{1}+z_{3})\Gamma(1-z_{1}-z_{3})}{\Gamma(2-z_{1}-z_{3}+n_3)\Gamma(1+z_1+z_2+z_3)\Gamma(1-z_1-z_2-z_3)} \nonumber \\ & \times \frac{\Gamma(1-z_1-z_2-z_3+n_3-n_2)\Gamma(1+z_{2}+z_{3})\Gamma(1-z_{2}-z_{3})}{\Gamma(2-z_{2}-z_{3}+n_1+n_3-n_2)}
    \end{align}
and the grouping of the singular factors in $\left( f_1 ,f_2, f_3 \right)$ is the same as for Type 2. However, we need to take into account the effect of the singular factor coming from the singular gamma function of the denominator, which gives
\begin{equation}
    \frac{z_1+z_2+z_3}{\left( -z_1, -z_2, (z_1+z_3)(z_2+z_3) \right)}= \frac{1}{\left( -1, -z_2, (z_1+z_3)(z_2+z_3) \right)}+ \frac{1}{\left( -z_1, -z_2, (z_1+z_3) \right)}
\end{equation}
The residue due to $\left( -1, -z_2, (z_1+z_3)(z_2+z_3) \right)$ is obviously zero. Therefore, the effective grouping of singular factors is $\left( -z_1, -z_2, z_1+z_3 \right) $. The corresponding contribution is then obtained by a nonresonant residue computation as for Type 1 and reads
\begin{align}
    S^{\text{unit}}_{2,3}= 
    \frac{1}{w^2}\sum_{n_1,n_2,n_3=0}^{\infty}\frac{{\Gamma}^{2}(1+n_3)\Gamma(1+n_1+n_2)\Gamma(2+n_1+n_2+n_3)}{\Gamma(2+n_1+n_3)\Gamma(2+n_2+n_3)} \frac{(u/w)^{n_1}(v/w)^{n_2}(1/w)^{n_3}}{n_1!\,n_2!\, n_3! } 
\end{align}\\

      {\bf Set of poles II:} $(n_1,n_2,-1-n_2-n_3)$\\

      These are the poles coming from the 3-combination $(1,2,6)$.
      
    We shift the poles to the origin by the change of variable $z_i \to z_i+n_i$ for $i=1,2$ and $z_3\to z_3-1-n_2-n_3$. The integrand of \eqref{MBtriangle} becomes
    \begin{align}\label{Cone_21_Singular_2}
      & \left(-u\right)^{z_{1}+n_1} \left(-v\right)^{z_{2}+n_2}  \left(-w\right)^{z_{3}-1-n_2-n_3}
    \frac{\Gamma(-z_{1}-n_1)\Gamma(-z_{2}-n_2)\Gamma(-z_{3}+1+n_2+n_3)}{\Gamma(1+z_1+z_2+z_3+n_1-n_3)}
   \nonumber \\ & \times \Gamma(1+z_{1}+z_{2}+n_1+n_2){\Gamma}(z_{1}+z_{3}+n_1-n_2-n_3){\Gamma}(z_{2}+z_{3}-n_3)
    \end{align}
As mentioned above, the poles for which the fifth gamma function is singular shall be omitted because such poles overlap with poles of Set I that have already been considered before. Thus we have only one type of pole.\\

    $\bullet$ Type 1: $n_1\geq 1+n_2+n_3$ and remaining summation variables run from $0$ to $\infty$.\\\\ 
    For this type of poles, only the first, second and sixth gamma functions are singular at origin. Following the same approach as for Set I Type 1, one obtains
\begin{align}
    S^{\text{unit}}_{2,4}= 
    \frac{u}{w}\sum_{n_1,n_2,n_3=0}^{\infty}(-1)^{1+n_2}\frac{{\Gamma}^{2}(1+n_1)\Gamma(1+n_2+n_3)\Gamma(2+n_1+2n_2+n_3)}{\Gamma(2+n_1+n_2)\Gamma(2+n_1+n_2+n_3)} \frac{u^{n_1}(uv/w)^{n_2}(u/w)^{n_3}}{n_1!\,n_2!\, n_3! } 
\end{align}
The final series representation is then
\begin{equation}\label{S2unit}
    S^{\text{unit}}_2=S^{\text{unit}}_{2,1}+S^{\text{unit}}_{2,2}+S^{\text{unit}}_{2,3}+S^{\text{unit}}_{2,4}
\end{equation}
Due to symmetry of $H_C$ (which can be seen in the MB integral in Eq.(\ref{MBtriangle})), one can obtain the series representation $S^{\text{unit}}_{5}$ and $S^{\text{unit}}_{11}$ by appropriate transformations on $S^{\text{unit}}_{2}$.

Once again Eq.(\ref{Srivastava_continuation}) provides an analytic cross-check of our result. Indeed, by the substitution $(\alpha,\beta,\beta',\gamma) \to (1,1+\epsilon,1+\epsilon',2)$ into Eq.(\ref{Srivastava_continuation}) and by carefully taking the limit $\epsilon , \epsilon' \to 0$ on the RHS, one obtains an analytic matching with the expression of $S^{\text{unit}}_{2}$ given in Eq.(\ref{S2unit}) which therefore provides the expected analytic continuation of $S^{\text{unit}}_{1}$.

The other independent series representations are given in the Appendix, see Section \ref{UnitApp}.

\section{Conclusions\label{conclusions}}
In this paper we have presented the results for the analysis of
the massive one-loop 3- and $n$-point conformal Feynman integrals, which are nontrivial
cases for the illustration of a newly introduced method of
evaluation of $N$-fold MB integrals.  The method, that
was presented in \cite{ABFGgeneral}, is based on a study of conic hulls associated with the gamma functions in the MB integrand.  In the present work, we
provided a detailed application of that method, briefly described for completeness in Section \ref{Proof_conj}, on a family of MB
integrals that have sufficient
structure to help us illustrate some of the subtle aspects of the newly
introduced method. The analysis yields detailed 
checks on some results already available in the literature, and provides many new results. In particular, we have settled two conjectures based on the analysis of the objects above
from a Yangian point of view \cite{Loebbert:2020glj} and derived from them a set of new quadratic transformations for a certain class of multiple hypergeometric series. Furthermore, our analysis has shown that new results can be obtained
on hitherto unknown analytic continuations of the Srivastava's $H_C$ triple hypergeometric series.

Our work thus provides one more intimate link
between solutions of Feynman integrals and hypergeometric functions theory, by explicitly showing how the former can well become an interesting testing ground to find new results in the latter.

Here we briefly summarize what has been done in the foregoing.  In Section \ref{Proof_conj}, we have provided a short recapitulation of the MB computational method introduced in \cite{ABFGgeneral} and used the two sets of conformal variables chosen in \cite{Loebbert:2020glj} to derive the corresponding $n(n-1)/2$-fold MB representations for the massive one-loop conformal $n$-point integral. By using the method in \cite{ABFGgeneral}, we show that one series representation for each of these MB integrals, in the nonresonant case of generic propagator powers, consists of a single series: the one that corresponds to the trivial conic hull. The MB integrals are then solved to show explicitly that these single series representations are in agreement with the expressions given in the conjectures of \cite{Loebbert:2020glj}. In Sections \ref{MB_3point} and \ref{Non_Resonant}, the above integrals are solved for the specific case of $n=3$, still in the simpler nonresonant class of conformal integrals having generic powers of the propagators. For one set of conformal variables, the results are in the form of Srivastava's $H_C$ triple hypergeometric series and its analytic continuations, which converge in different kinematic regions. The other set provides alternative series representations, one of them leading to an interesting new quadratic transformation of $H_C$ in terms of another well-known Srivastava's triple hypergeometric series: $H_B$. This quadratic transformation can be extended to a certain class of multiple hypergeometric series, as it is conjectured in Section \ref{quadratic}.
 As an aside, we also demonstrate in Section \ref{Non_Resonant} a difference between the Yangian bootstrap approach and our MB method, in that the latter is able to easily distinguish the spurious zeros of the former, that do not contribute to the series solutions, and which therefore need to be discarded.  In Section \ref{resonant}, the MB representation of the 3-point integral of Section \ref{MB_3point} associated with Set 1 is evaluated for unit propagator powers. This is a resonant case, more difficult to compute, which shows another powerful ability of our MB computational method. The results of Section \ref{Non_Resonant} and \ref{resonant} are partly given in the Appendix.

The main results presented in this paper are a consequence of the remarkable first
solution to the general problem of finding series representations to a general
$N$-fold MB integral \cite{ABFGgeneral}.  
The future is rich with possibilities for the exploration of their properties.

\bigskip

\bigskip

\noindent {\bf Acknowledgements}

\vspace{0.5cm}

We thank F.~Loebbert, J.~Miczajka, D.~M\"uller and H.~M\"unkler for correspondence about some results of  \cite{Loebbert:2020hxk}, Apoorva D. Patel for enlightening discussions, and Souvik Bera and Tanay Pathak for their help with the convergence analysis.
S. G. thanks Collaborative Research
Center CRC 110 Symmetries and the Emergence of
Structure in QCD for supporting the research through grants.

\bigskip

\bigskip

\section{Appendix}

\subsection{Set 1:  Building blocks\label{BBset1}}

\begin{align}
    B_{1,2,3}= 
    \sum_{n_1,n_2,n_3=0}^{\infty}\frac{\Gamma(a_1+n_1+n_2)\Gamma(a_2+n_1+n_3)\Gamma(a_3+n_2+n_3)}{\Gamma(\frac{1+a_1+a_2+a_3}{2}+n_1+n_2+n_3)} \frac{u^{n_1}v^{n_2}w^{n_3}}{n_1!\,n_2!\, n_3! } 
    \end{align}
    \begin{align}
    B_{1,2,5}=(-w)^{-a_2} 
    &\sum_{n_1,n_2,n_3=0}^{\infty}\frac{\Gamma(a_1+n_1+n_2)\Gamma(-a_2+a_3+n_2-n_1-n_3)}{\Gamma(\frac{1+a_1-a_2+a_3}{2}+n_2-n_3)} \nonumber \\ & \times \Gamma(a_2+n_1+n_3) \frac{\left(-u/w\right)^{n_1}v^{n_2}\left(1/w\right)^{n_3}}{n_1!\,n_2!\, n_3! } 
\end{align}
\begin{align}
    B_{1,2,6}= (-w)^{-a_3}
    &\sum_{n_1,n_2,n_3=0}^{\infty}\frac{\Gamma(a_1+n_1+n_2)\Gamma(a_2-a_3+n_1-n_2-n_3)}{\Gamma(\frac{1+a_1+a_2-a_3}{2}+n_1-n_3)} \nonumber \\ & \times \Gamma(a_3+n_2+n_3)\frac{u^{n_1}(-v/w)^{n_2}(1/w)^{n_3}}{n_1!\,n_2!\, n_3! } 
\end{align}
$B_{1,2,6}$ can be obtained by applying the transformation $\left(u, a_{2}\right) \leftrightarrow \left( v, a_{3}\right)$ on $B_{1,2,5}$.
\begin{align}
    B_{2,4,6}= 
    (-u)^{-a_1}(-w)^{-a_3}&\sum_{n_1,n_2,n_3=0}^{\infty}\frac{\Gamma(a_1+n_1+n_2)\Gamma(-a_1+a_2-a_3-2n_1-n_2-n_3)}{\Gamma(\frac{1-a_1+a_2-a_3}{2}-n_1-n_2-n_3)} \nonumber \\ & \times \Gamma(a_3+n_1+n_3) \frac{(v/(u \, w))^{n_1}(1/u)^{n_2}(1/w)^{n_3}}{n_1!\,n_2!\, n_3! } 
\end{align}
\begin{align}
    B_{2,5,6}= (-u)^{-a_2+a_3}(-w)^{-a_3}
    &\sum_{n_1,n_2,n_3=0}^{\infty}\frac{\Gamma(a_2-a_3-n_1+n_2-n_3)\Gamma(a_3+n_1+n_3) }{\Gamma(\frac{1+a_1-a_2+a_3}{2}+n_1-n_2)} \nonumber \\ & \times \Gamma(a_1-a_2+a_3+2n_1-n_2+n_3) \, \frac{(u\,v/w)^{n_1} (1/u)^{n_2} (-u/w)^{n_3}}{n_1!\,n_2!\, n_3! } 
\end{align}
\begin{align}
B_{1,4,5}=(-v)^{-a_1}(-w)^{-a_2}&\sum_{n_1,n_2,n_3=0}^{\infty}  \frac{\Gamma(a_1+n_1+n_2)\Gamma(a_3-a_1-a_2-2n_1-n_2-n_3)}{\Gamma(\frac{1-a_1-a_2+a_3}{2}-n_1-n_2-n_3)} \nonumber \\ & \times \Gamma(a_2+n_1+n_3)\frac{(u/(v \, w))^{n_1} (1/v)^{n_2} (1/w)^{n_3}}{n_1! \, n_2! \, n_3!}
\end{align}
$B_{1,4,5}$ can be obtained by applying the transformation $\left(u, a_{2}\right) \leftrightarrow \left( v, a_{3}\right)$ on $B_{2,4,6}$.
\begin{align}
B_{1,4,6}=(-v)^{-a_1}(-w)^{a_1-a_3}&\sum_{n_1,n_2,n_3=0}^{\infty} \frac{\Gamma(-a_1+a_3-n_1-n_2+n_3)\Gamma(a_1+n_1+n_2)}{\Gamma(\frac{1+a_1+a_2-a_3}{2}+n_1-n_3)} \nonumber \\ & \times \Gamma(a_1+a_2-a_3+2n_1+n_2-n_3)\frac{(u \, w/v)^{n_1}(-w/v)^{n_2}(1/w)^{n_3}}{n_1! \, n_2! \, n_3!}
\end{align}
$B_{1,4,6}$ can be obtained by applying the transformation $\left(u, v, w, a_{1}, a_{2}, a_{3}\right) \rightarrow\left(w, u, v, a_{2}, a_{3}, a_{1}\right)$ on $B_{2,5,6}$.
\begin{align}
B_{3,4,6}=(-u)^{-a_1+a_3}(-v)^{-a_3}&\sum_{n_1,n_2,n_3=0}^{\infty} \frac{\Gamma(a_1-a_3-n_1+n_2-n_3)\Gamma(a_3+n_1+n_3)}{\Gamma(\frac{1-a_1+a_2+a_3}{2}+n_1-n_2)} \nonumber \\ & \times\Gamma(-a_1+a_2+a_3+2n_1-n_2+n_3) \frac{(u \, w/ v)^{n_1} (1/u)^{n_2} (-u/v)^{n_3}}{n_1! \, n_2! \, n_3!}
\end{align}
$B_{3,4,6}$ can be obtained by applying the transformation $\left(v, a_{1}\right) \leftrightarrow\left(w, a_{2}\right)$ on $B_{2,5,6}$.
\begin{align}
B_{3,5,6}=(-u)^{-a_2} (-v)^ {-a_3} &\sum_{n_1,n_2,n_3=0}^{\infty} \frac{\Gamma(a_2+n_1+n_2)\Gamma(a_1-a_2-a_3-2n_1-n_2-n_3)}{\Gamma(\frac{1+a_1-a_2-a_3}{2}-n_1-n_2-n_3)} \nonumber \\ & \times\Gamma(a_3+n_1+n_3) \frac{(w/(u \, v))^{n_1} (1/u)^{n_2}(1/v)^{n_3}}{n_1! \, n_2! \, n_3!}
\end{align}
$B_{3,5,6}$ can be obtained by applying the transformation $\left(v, a_{1}\right) \leftrightarrow\left(w, a_{2}\right)$ on $B_{2,4,6}$.
\begin{align}
    B_{4,5,6}&= i^{-a_1-a_2-a_3} ( \sqrt{u})^{-a_1-a_2+a_3}( \sqrt{v})^{-a_1+a_2-a_3}( \sqrt{w})^{a_1-a_2-a_3}\nonumber \\ & \times
    \sum_{n_1,n_2,n_3=0}^{\infty}\dfrac{\Gamma\left(\frac{a_1+a_2-a_3+n_1+n_2-n_3}{2}\right)\Gamma\left(\frac{a_1-a_2+a_3+n_1-n_2+n_3}{2}\right)\Gamma\left(\frac{-a_1+a_2+a_3-n_1+n_2+n_3}{2}\right)}{\Gamma\left(\frac{1-n_1-n_2-n_3}{2}\right)} \nonumber \\ & \times  \frac{\left(\sqrt{-w/(u \, v)}\right)^{n_1}\left(\sqrt{-v/(u \, w)}\right)^{n_2}\left(\sqrt{-u/(v \, w)}\right)^{n_3}}{2\,\,n_1!\,n_2!\, n_3! } 
\end{align}

We end this section by noting that only 5 out of the 17 building blocks are independent as the remaining 12 can be derived by applying appropriate transformations on them. This is a manifestation of the symmetric structure of the MB integrand in \eqref{MBtriangle}  and more generally of the $H_C$ triple hypergeometric function.

\subsection{Set 1: Convergence regions\label{conv_set1}}

Based on the master series conjecture \cite{ABFGgeneral} and applying Horn's theorem we obtain the convergence regions for the series representations of Set 1 except for $S_5$ for which we give only a conjecture of its expression. For $S_1,..., S_4$ we also computed the convergence regions of each of the building blocks and, using these, found agreement with those obtained from the master series, thereby, reconfirming the conjecture.\\

\begin{figure}[h] \label{Old_MB_ROC}
\centering
\includegraphics[width=11.6cm]{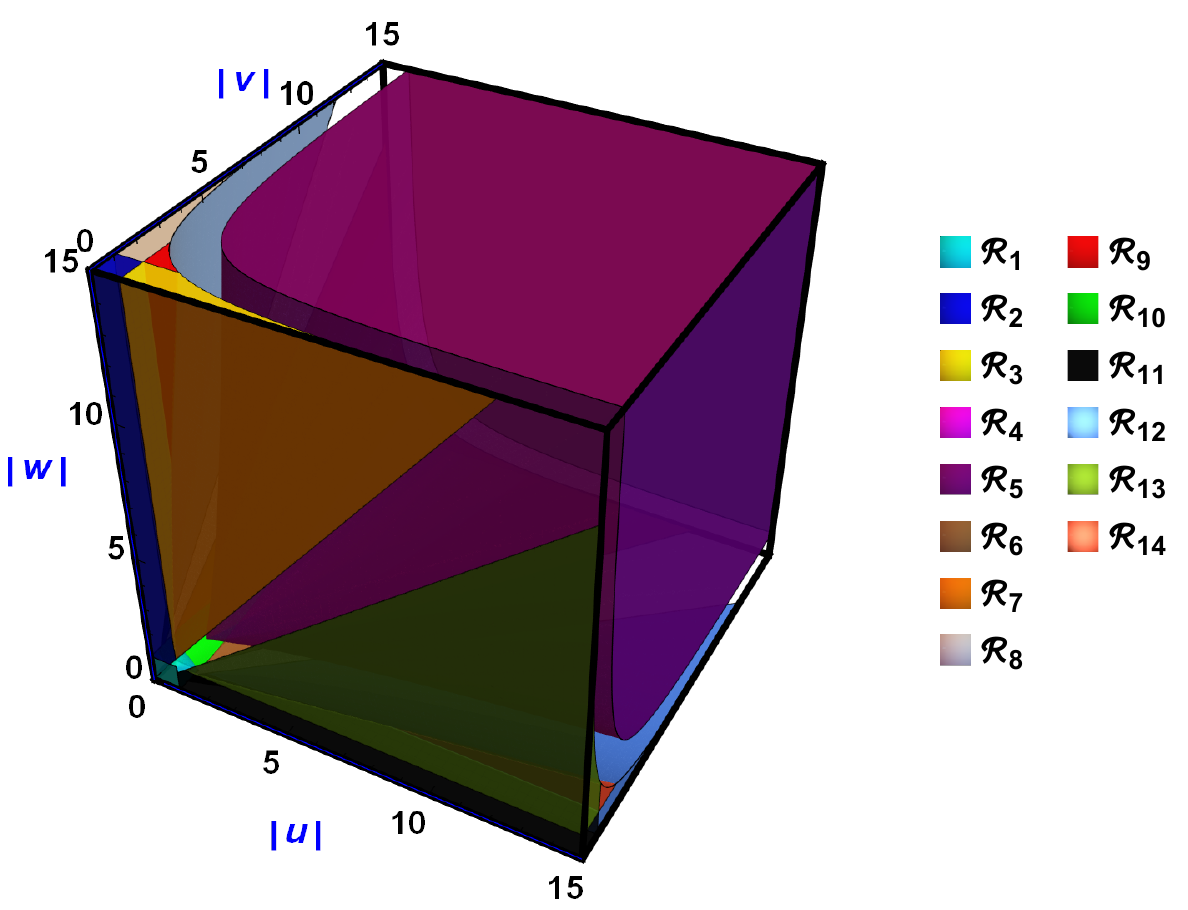}
\caption{ Convergence regions of the 14 series representations of the MB integral $J_1$ in Eq.\eqref{MBtriangle}.\label{FigConvSet1}}
\end{figure}

The convergence region of  $S_1$ comes from the well-known convergence properties of $H_C$ which give
    \begin{equation}\label{R1set1}
        \bm{\mathcal{R}_{1}}=\left\{ \left|u\right|<1\,\,\cap\,\,\left|v\right|<1\,\,\cap\,\,\left|w\right|<1\,\,\cap\,\, \left|u\right|+\left|v\right|+\left|w\right|<2+2\sqrt{(1-\left|u\right|)(1-\left|v\right|)(1-\left|w\right|)} \right\}
    \end{equation}
This is the cyan region in Figure \ref{FigConvSet1}.

\medskip

\textit{Convergence region of $S_2$:} the characteristic list of the master series due to poles at $(n_1,n_2,-n_3)$ is $\{n_1+n_2,n_1-n_3,n_2-n_3,n_3-n_1-n_2,n_3,n_3\}$. Applying Horn's theorem yields the following convergence region:
    \begin{equation}
        \bm{\mathcal{R}_{2}}=\left\{ \left|u\right|<1\,\,\cap\,\,\left|v\right|<1\,\,\cap\,\,\left|w\right|>1\,\,\cap\,\, \left|u\right|< \left|v\right|+\left|w\right|+2\left|v\right|\left|w\right|-2\sqrt{\left|v\right|\left|w\right|(1+\left|v\right|)(1+\left|w\right|)} \right\}
    \end{equation}
which is the blue region in Figure \ref{FigConvSet1}.

\medskip

\textit{Convergence region of  $S_3$:} three conic hulls intersect for this series representation and the corresponding master series has the characteristic list  $\{n_1+n_2,n_1-n_3,n_2-n_3,n_3-n_1-n_2,n_3,n_3\}$. From Horn's theorem one then obtains
    \begin{align}
        \bm{\mathcal{R}_{3}}=\Bigg\{ & \left|u\right|>1\,\,\cap\,\,\left|v\right|<1\,\,\cap\,\,\left|\frac{u}{w}\right|<1\,\,\cap\,\, \left| \frac{4 u\, v}{w}\right|< \left(\left| \frac{u}{w}\right| -1\right)^{2} \,\,\cap\,\, \left| v \right|< \left| w \right|+\left| u \right|+2 \sqrt{\left| u \, w\right|(1-\left|v\right|)} \nonumber \\& \cap\,\, \left| v \right|< \left| w \right|+\left| u \right|-2 \sqrt{\left| u \, w\right|(1+\left|v\right|)} \,\,\cap\,\, \left| v \right|< \left| w \right|+\left| u \right|+2 \sqrt{\left| u \, w\right|(1+\left|v\right|)}  \Bigg\}
    \end{align}
which is the yellow region in Figure \ref{FigConvSet1}.

\medskip

\textit{Convergence region of  $S_4$:} in this case, four conic hulls intersect and the corresponding master series has the characteristic list 
    $\{n_1 - n_2, n_2, -n_2 + n_3, -n_1 + 2 n_2 - n_3, n_1 - n_2 + n_3, n_2\}$. Applying Horn's theorem yields 
\begin{align}
        \bm{\mathcal{R}_{4}}&=\Bigg\{  \left|u\right|>1\,\,\cap\,\,4\left|\frac{u w}{v}\right|<1\,\,\cap\,\,\left|w\right|>1\,\,\cap\,\, 
        |u|+|w|<|v|+2\sqrt{|u||v|(|w|-1)}\nonumber \\& \,\,\cap\,\, |u|+|v|<|w|+2\sqrt{|u||v|(|w|+1)} \,\,\cap\,\, \sqrt{\left|\frac{v}{u w}\right|}>1+ \sqrt{1+\frac{1}{\left|u\right|}}\,\,\cap\,\, \sqrt{\left|\frac{u}{v}\right|}<\sqrt{1-2\sqrt{\left|\frac{u w}{v}\right|}} \Bigg\}
    \end{align}
which is the pink region in Figure \ref{FigConvSet1}.

\medskip

 \textit{Convergence region of  $S_5$:} here, four conic hulls intersect and the corresponding master series has the characteristic list 
    $\{\frac{n_1+n_2-n_3}{2},\frac{n_1-n_2+n_3}{2},\frac{-n_1+n_2-n_3}{2},\frac{n_1+n_2+n_3}{2}\}$. This characteristic list matches with the characteristic list of the building block $B_{4,5,6}$. We were unable to find its convergence region using Horn's theorem, due to computational complexity. Therefore, we conjecture the convergence region of $S_5$ as the intersection of the convergence regions of the remaining three building blocks $B_{1,4,5}, B_{2,4,6}$ and $B_{3,5,6}$, which are relatively easier to calculate and guarantees that the actual convergence region cannot be larger.
    \begin{equation}
        \bm{\mathcal{R}_5}= R_{1,4,5} \cap R_{2,4,6} \cap R_{3,5,6}
    \end{equation}
    where,
    \begin{align}
        R_{1,4,5}&=\Bigg\{\frac{1}{| w| }<1 \, \cap \, \sqrt{\left| \frac{u}{v w}\right| }<2\cap \sqrt{| v| }<| v| \, \cap \, \frac{1}{| v| }<1\cap \sqrt{| w| }<| w| \nonumber \\ & \cap
   \left(\sqrt{\left| \frac{u}{v w}\right| }<1\cup \left(\frac{1}{\sqrt{| v| }}<\sqrt{2-\sqrt{\left| \frac{u}{v w}\right| }} \sqrt[4]{\left|
   \frac{u}{v w}\right| }\nonumber \cap \frac{1}{\sqrt{| w| }}<\sqrt{2-\sqrt{\left| \frac{u}{v w}\right| }} \sqrt[4]{\left| \frac{u}{v w}\right|
   }\right)\right) \nonumber \\ & \cap \, \, \left| \frac{u}{v w}\right| +\frac{1}{| v| }+\frac{1}{| w| }<2 \sqrt{\left(\frac{1}{| v| }-1\right) \left(\frac{1}{| w|
   }-1\right)}+2 \Bigg\}
    \end{align}
    and, $R_{2,4,6}$ and $R_{3,5,6}$ can be obtained from $R_{1,4,5}$ by the transformations $(u,v)\leftrightarrow(v,u)$ and $(u,v,w)\rightarrow(w,u,v)$, respectively. 
    
    $\bm{\mathcal{R}_5}$ is the purple region in Figure \ref{FigConvSet1}.
\subsection{Set 2: building blocks\label{BBset2}}

The integral $J_2$ in Eq.\eqref{Conformal_Triangle_2} has 14 series representations, which for the nonresonant generic propagator power case are built from 17 building blocks. \\
As explained in Section \ref{Series_set2}, we give below five series representation only, as the remaining nine series representations can be obtained by exploiting the symmetry of the MB integrand in Eq.(\ref{Conformal_Triangle_2}).
\begin{enumerate}
    \item $S'_1$: It is equal to the building block $B'_{1,2,3}$.
    \begin{align} \label{New_arbitrary_propagator_1}
        B'_{1,2,3} = \sum_{n_1,n_2,n_3=0}^{\infty}\Gamma \left( \frac{a_1+n_1+n_2}{2}\right)
   &\Gamma \left( \frac{a_2+n_1+n_3}{2}\right)\Gamma \left( \frac{a_3+n_2+n_3}{2}\right)\nonumber \\ 
    & \times \frac{(-2u')^{n_1} (-2v')^{n_2} (-2w')^{n_3}}{n_1! \, n_2! \, n_3!}	
    \end{align}
    where, $u' , v' , w' $ are the new conformal variables in Section 3.2.
    \item $S'_2$: It is equal to the sum of building blocks $B'_{1,2,5}$ and $B'_{1,2,6}$.
    \begin{align} \label{New_arbitrary_propagator_2}
        B'_{1,2,5}= 2 (2 w')^{-a_2} \sum_{n_1,n_2,n_3=0}^{\infty}
& \Gamma \left( \frac{a_1+n_1+n_2}{2}\right) \Gamma \left(\frac{-a_2+a_3-n_1+n_2-2n_3}{2}\right) \nonumber \\ 
    & \times \Gamma \left( a_2+n_1+2n_3\right)\frac{(-u'/w')^{n_1} (-2v')^{n_2} (-1/(4{w'}^2))^{n_3}}{n_1! \,n_2! \, n_3!}   
    \end{align}
and $B'_{1,2,6}$ can be obtained from $B'_{1,2,5}$ by the transformation $(u',a_2) \leftrightarrow (v',a_3)$.
    \item $S'_3$: It is equal to the sum of building blocks $B'_{1,2,5}$, $B'_{2,4,6}$ and $B'_{2,5,6}$. 
    \begin{align}  \label{New_arbitrary_propagator_12}
  B'_{2,4,6} &= 4 (2u')^{-a_1}(2w')^{-a_3} \sum_{n_1,n_2,n_3=0}^{\infty}\Gamma \left(a_1+n_1+2n_2\right)
    \Gamma\left(a_3+n_1+2n_3\right)\nonumber \\ 
    & \times 
   \Gamma \left( \frac{-a_1+a_2-a_3}{2}-n_1-n_2-n_3\right)\frac{(-v'/(2 u' \, w'))^{n_1}(-1/(4{u'}^2))^{n_2}(-1/(4{w'}^2))^{n_3}}{n_1! \, n_2! \, n_3!} 
   \end{align}
   \begin{align}  \label{New_arbitrary_propagator_13}
   B'_{2,5,6} &= 4 (2 u')^{-a_2+a_3}(2w')^{-a_3}\sum_{n_1,n_2,n_3=0}^{\infty}\Gamma \left(a_2-a_3-n_1+2n_2-2n_3\right)
    \Gamma\left(a_3+n_1+2n_3\right)\nonumber \\ 
    & \times 
    \Gamma \left( \frac{a_1-a_2+a_3}{2}+n_1-n_2+n_3\right)\frac{(-2u'\,v'/w')^{n_1} (-1/(4{u'}^2))^{n_2}(-{u'}^2/{w'}^2)^{n_3}}{n_1! \, n_2! \, n_3!}
   \end{align}
and $B'_{1,2,5}$ is given in \eqref{New_arbitrary_propagator_2}.
\item $S'_4$: It is equal to the sum of building blocks $B'_{1,4,5}$, $B'_{1,4,6}$, $B'_{3,4,6}$ and $B'_{3,5,6}$. $B'_{1,4,5}$ can be obtained from $B'_{2,4,6}$ by the transformation $(a_2,u') \leftrightarrow (a_3,v')$. $B'_{1,4,6}$ can be obtained from $B'_{2,5,6}$ by the transformation $(a_1,a_2,a_3,u',v',w') \to (a_2,a_3,a_1,w',u',v')$. $B'_{3,4,6}$ and $B'_{3,5,6}$ can be obtained by the transformation $(a_1,v') \leftrightarrow (a_2,w')$ on $B'_{2,5,6}$ and $B'_{2,4,6}$, respectively.
\item $S'_5$: It is equal to the sum of building blocks $B'_{1,4,5}$, $B'_{2,4,6}$, $B'_{3,5,6}$ and $B'_{4,5,6}$. 
\begin{align}
   B'_{4,5,6} &= 4 (\sqrt{2 u'})^{-a_1-a_2+a_3}(\sqrt{2 v'})^{-a_1+a_2-a_3}
   (\sqrt{2 w'})^{a_1-a_2-a_3} \nonumber \\ & \times
   \sum_{n_1,n_2,n_3=0}^{\infty} \Gamma \left(\frac{a_1-a_2+a_3}{2}+n_1-n_2+n_3\right) \Gamma
   \left(\frac{-a_1+a_2+a_3}{2}-n_1+n_2+n_3\right) \nonumber \\
    & \times 
  \Gamma \left(\frac{a_1+a_2-a_3}{2}+n_1+n_2-n_3\right)  \frac{(-w' /( 2 u'\, v'))^{n_1}(-v' /( 2 u'\, w'))^{n_2}(-u' /( 2 v'\, w'))^{n_3} }{n_1! \, n_2! \, n_3!}   
   \end{align}
   and the remaining building blocks are already known.
\end{enumerate}

\subsection{Set 2: Convergence regions\label{conv_set2}}

Using the master series conjecture \cite{ABFGgeneral} and applying Horn's theorem we obtain the convergence regions for the series representations of Set 2. The difference with Set 1 is that all convergence regions have been obtained in this way. We also computed the convergence regions of each of the building blocks and, using these, found agreement with those obtained from the master series, thereby reconfirming the conjecture in \cite{ABFGgeneral}.\\

   $\bullet$ $S'_1$: The set of poles for the master series is $(n_1,n_2,n_3)$, and its characteristic list is $\{ \frac{n_1+n_2}{2},\frac{n_1+n_3}{2},\frac{n_2+n_3}{2} \}$.
    \begin{equation}
        \bm{\mathcal{R}'_{1}}=| x \, y \, z | +| x| ^2+| y| ^2+| z| ^2<4
    \end{equation}
where $x=2-4u$, $y=2-4v$ and $z=2-4w$; we use this notation throughout this subsection.\\

    $\bullet$ $S'_2$: The set of poles for the master series is $(n_1,n_2,-n_3)$, and its characteristic list is $\{n_3,n_3, \frac{n_1+n_2}{2},\frac{n_1-n_3}{2},\frac{n_2-n_3}{2} \}$. 
    \begin{align}
        \bm{\mathcal{R}'_{2}} =  &\frac{| x| ^2+4}{| z| ^2}<1\cap \frac{| y| ^2+4}{| z| ^2}<1 \cap \left(| y| ^2+2\right) | z| ^2> \left(| y|  | z|  \sqrt{\left(| y|
   ^2+4\right) \left(| z| ^2-4\right)}+2 |x|^2+2 |y|^2+8\right) \nonumber \\ & \cap | y|  | z|  \sqrt{\left(| y| ^2+4\right) \left(| z| ^2+4\right)}+2 |x|^2<|z|^2
   \left(| y| ^2+2\right)+2 |y|^2+8\cap | y| <2\cap | x| ^2+| y| ^2<4
    \end{align}\\
    
$\bullet$ $S'_3$: The set of poles for the master series is $(n_3-n_1,n_2,-n_3)$, and its characteristic list is $\{n_1-n_3,n_1,n_3,n_3, \frac{n_3+n_2-n_1}{2},\frac{-n_1}{2},\frac{n_2-n_3}{2} \}$.
    \begin{equation}
        \bm{\mathcal{R}'_{3}}=| y| <2\,\, \cap \,\, | x \, y \, z| +| y| ^2+|x|^2+4<|z|^2\,\, \cap \,\, | y| ^2+4<|x|^2
    \end{equation}\\
    
$\bullet$ $S'_4$: The set of poles for the master series is $(n_2-n_1,-n_2,n_2-n_3)$, and its characteristic list is $\{n_1-n_2,n_1,n_2,n_2,n_3,n_3-n_2, \frac{-n_1}{2},\frac{2n_2-n_1-n_3}{2},\frac{-n_3}{2} \}$.
    \begin{align}
        \bm{\mathcal{R}'_{4}}=| x| >2\,\, \cap \,\, & | x \, y \, z| +| x| ^2+| z| ^2+4< |y|^2\,\, \cap \,\, \left(| x| +\sqrt{| x| ^2+4}\right) | z|<2 | y|\,\, \nonumber \\ & \hspace{2cm}\cap \,\,
   \left(| x| -\sqrt{| y| } \sqrt{| y| -| x|  | z| }\right) | x \, y \, z| <0\,\, \cap \,\, | z| >2
    \end{align}\\
    
\begin{figure}[h]
\centering
\includegraphics[width=11.6cm]{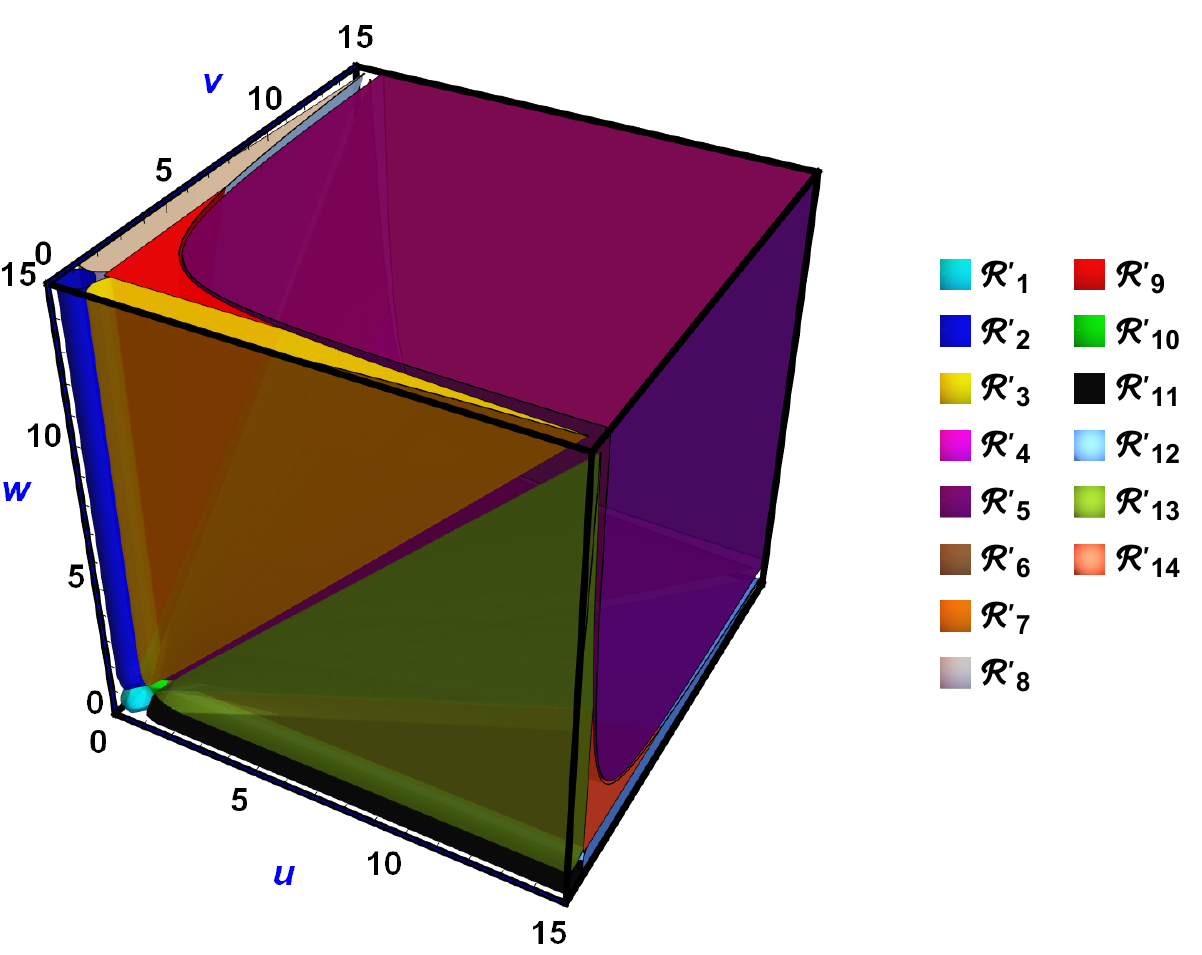}
\caption{ Convergence regions of the 14 series representations of the MB integral $J_2$ in Eq.(\ref{Conformal_Triangle_2}), for positive values of $u,v$ and $w$.}
\label{New_MB_ROC}
\end{figure}

$\bullet$ $S'_5$: The set of poles for the master series is $(-n_1-n_2+n_3,-n_1+n_2-n_3,n_1-n_2-n_3)$, and its characteristic list is $\{ n_1+n_2-n_3,n_1-n_2+n_3,-n_1+n_2+n_3 \}$.
    \begin{align}
        \bm{\mathcal{R}'_5}= | x| ^2+| y| ^2+| z| ^2 < 4 +| x \, y \, z| \,\, \cap \,\, \left| y\right|^2 +\left| z\right|^2 < \left| x \, y \, z\right|
    \end{align}
The convergence regions of the remaining 9 series representation can be obtained from the above regions, thanks to the symmetry of the MB integral in \eqref{Conformal_Triangle_2}. The convergence regions of all the 14 series representations are displayed in Figure \ref{New_MB_ROC}.

\subsection{Set 1: unit propagator powers\label{UnitApp}}

The series representation $S^{\text{unit}}_3$  is given by
\begin{align}
S^{\text{unit}}_3=S^{\text{unit}}_{3,1}+S^{\text{unit}}_{3,2}+S^{\text{unit}}_{3,3}
\end{align}
where
\begin{align}
    S^{\text{unit}}_{3,1}= -\frac{v}{w}
    \sum_{n_1,n_2,n_3=0}^{\infty}\frac{{\Gamma}^{2}(1+n_2)\Gamma(1+n_1+n_3)\Gamma(2+2n_1+n_2+n_3)}{\Gamma(2+n_1+n_2)\Gamma(2+n_1+n_2+n_3)} \frac{\left(-{u\, v}/{w}\right)^{n_1}\left(v\right)^{n_2}\left({v}/{w}\right)^{n_3}}{n_1!\,n_2!\, n_3! } 
\end{align}
this is same as $S^{\text{unit}}_{2,1}$.
\begin{align}
    S^{\text{unit}}_{3,2}= & \frac{1}{w}
    \sum_{n_1,n_2,n_3=0}^{\infty}\frac{{\Gamma}(1+n_1+n_2+n_3)\Gamma(1+n_1+2n_2+n_3)}{\Gamma(1+n_1+n_2)\Gamma(1+n_2+n_3)} \bigg( \text{log} (-u)-\text{log} (-w) -\psi(1+n_1)\nonumber \\ &  -\psi(1+n_1+n_2)+\psi(1+n_1+n_2+n_3)+\psi(1+n_1+2n_2+n_3) \bigg) \frac{(u/w)^{n_1}(-u \, v/w)^{n_2}(v/w)^{n_3}}{n_1!\,n_2!\, n_3! } 
\end{align}
\begin{align}
    S^{\text{unit}}_{3,3}=-\frac{1}{u\,w} 
    \sum_{n_1,n_2,n_3=0}^{\infty}\frac{{\Gamma}(1+n_1+n_2)\Gamma(1+n_1+n_3)\Gamma(1+n_1+n_2+n_3)}{\Gamma(2+2n_1+n_2+n_3)} \frac{(-v/(u\,w))^{n_1}(1/u)^{n_2}(1/w)^{n_3}}{n_1!\,n_2!\, n_3! } 
\end{align} 
The series representation $S^{\text{unit}}_4$  is given by
\begin{align}
S^{\text{unit}}_4=&\bigg(S^{\text{unit}}_{4,1}+ (u, v, w) \rightarrow(w, u, v)\bigg)+ S^{\text{unit}}_{4,2}+S^{\text{unit}}_{4,3} 
\end{align}
where
\begin{align}
    S^{\text{unit}}_{4,1}=-\frac{1}{v \, w} 
    \sum_{n_1,n_2,n_3=0}^{\infty}\frac{{\Gamma}(1+n_1+n_2)\Gamma(1+n_1+n_3)\Gamma(1+n_1+n_2+n_3)}{\Gamma(2+2n_1+n_2+n_3)} \frac{(- u / (v \, w))^{n_1}(1/v)^{n_2}(1/w)^{n_3}}{n_1!\,n_2!\, n_3! } 
\end{align}  
\begin{align}
    S^{\text{unit}}_{4,2}= &\frac{1}{v} 
    \sum_{n_1,n_2,n_3=0}^{\infty}\frac{{\Gamma}(1+n_1+n_2+n_3)\Gamma(1+2n_1+n_2+n_3)}{\Gamma(1+n_1+n_2)\Gamma(1+n_1+n_3)} \bigg( \text{log} (-u)   \nonumber \\ & -\text{log} (-v)+\text{log} (-w)-\psi(1+n_1)-\psi(1+n_1+n_2)-\psi(1+n_1+n_3) \nonumber \\ &
   +\psi(1+n_1+n_2+n_3) +2\psi(1+2n_1+n_2+n_3) \bigg) \frac{(- u \,w /v)^{n_1}(w/v)^{n_2}(u/v)^{n_3}}{n_1!\,n_2!\, n_3! } 
\end{align}
\begin{align}
    S^{\text{unit}}_{4,3}= -\frac{1}{v^2}
    \sum_{n_1,n_2,n_3=0}^{\infty}\frac{{\Gamma}^{2}(1+n_3)\Gamma(1+n_1+n_2)\Gamma(2+n_1+n_2+n_3)}{\Gamma(2+n_1+n_3)\Gamma(2+n_2+n_3)} \frac{(u/v)^{n_1}(w/v)^{n_2}(1/v)^{n_3}}{n_1!\,n_2!\, n_3! } 
\end{align} 

The series representation $S^{\text{unit}}_5$ is given by:
\begin{align}
S^{\text{unit}}_5=&\bigg(S^{\text{unit}}_{5,1}+ (u, v, w) \rightarrow(v, u, w)+(u, v, w) \rightarrow(w, u, v) \bigg)+ \bigg(S^{\text{unit}}_{5,2}+ (n_1, n_2) \rightarrow\left(n_1+1/2,n_2+1/2\right) \nonumber \\ & + (n_1, n_3) \rightarrow\left(n_1+1/2,n_3+1/2\right) + (n_2, n_3) \rightarrow\left(n_2+1/2,n_3+1/2\right)  \bigg)
\end{align}
where
\begin{align}
    S^{\text{unit}}_{5,1}= -\frac{1}{v \, w}
    \sum_{n_1,n_2,n_3=0}^{\infty}\frac{{\Gamma}(1+n_1+n_2)\Gamma(1+n_1+n_3)\Gamma(1+n_1+n_2+n_3)}{2\,\Gamma(2+2n_1+n_2+n_3)} \frac{(-u/(v\,w))^{n_1}(1/v)^{n_2}(1/w)^{n_3}}{n_1!\,n_2!\, n_3! } 
\end{align}
\begin{align}
    S^{\text{unit}}_{5,2}= &\frac{i \pi^{3/2}}{\sqrt{u \, v \, w}}
    \sum_{n_1,n_2,n_3=0}^{\infty} \frac{\Gamma(1/2+n_1+n_2-n_3)\Gamma(1/2+n_1-n_2+n_3)}{2^{2n_1+2n_2+2n_3+1}\Gamma(1/2-n_1-n_2-n_3)} \nonumber\\ & \times \frac{\Gamma(1/2-n_1+n_2+n_3)}{\Gamma(n_1+1/2)\Gamma(n_2+1/2)\Gamma(n_3+1/2)} \frac{(-w/(u \, v))^{n_1}(-v/(u \, w))^{n_2}(-u/(v \, w))^{n_3}}{n_1!\,n_2!\, n_3! } 
\end{align}


\bigskip

\bigskip








\begin{thebibliography}{99}


\bibitem{ABFGgeneral}
B.~Ananthanarayan, S.~Banik, S.~Friot and S.~Ghosh, \textit{``Multiple Series Representations of $N$-fold Mellin-Barnes Integrals''} [arXiv:2012.15108 [hep-th]].

\bibitem{Loebbert:2020hxk}
F.~Loebbert, J.~Miczajka, D.~M\"uller and H.~M\"unkler,
Phys. Rev. Lett. \textbf{125} (2020) no.9, 091602
doi:10.1103/PhysRevLett.125.091602
[arXiv:2005.01735 [hep-th]].

\bibitem{Loebbert:2020glj}
F.~Loebbert, J.~Miczajka, D.~M\"uller and H.~M\"unkler,
[arXiv:2010.08552 [hep-th]].
 
\bibitem{Srivastava67}
H.~M.~Srivastava, Rend. Circ. Mat. Palermo (2) 16 (1967), 99-115. 



\bibitem{Srivastava}
H.~M.~Srivastava and P.~W.~Karlsson \textit{``Multiple gaussian hypergeometric series''}, Halsted Press (Ellis Horwood Limited, Chichester) (John Wiley and Sons, New York, 1985). 
 
\bibitem{Srivastava72}
H.~M.~Srivastava, Mat. Vesnik 9 (24) (1972), 101-107.   
  
 
  
  
\bibitem{Loebbert:2019vcj}
F.~Loebbert, D.~M\"uller and H.~M\"unkler,
Phys. Rev. D \textbf{101} (2020) no.6, 066006
doi:10.1103/PhysRevD.101.066006
[arXiv:1912.05561 [hep-th]].
  
\bibitem{Ananthanarayan:2020ncn}
B.~Ananthanarayan, S.~Banik, S.~Friot and S.~Ghosh,
Phys. Rev. D \textbf{102} (2020) no.9, 091901
doi:10.1103/PhysRevD.102.091901
[arXiv:2007.08360 [hep-th]].
  
  
 
\bibitem{Srivastava64}
H.~M.~Srivastava, Ganita 15 (1964), 97-108. 
  
  
  
\bibitem{Ananthanarayan:2019icl}
B.~Ananthanarayan, S.~Friot and S.~Ghosh,
Eur. Phys. J. C \textbf{80} (2020) no.7, 606
doi:10.1140/epjc/s10052-020-8131-3
[arXiv:1911.10096 [hep-ph]].

\bibitem{Ananthanarayan:2020acj}
B.~Ananthanarayan, S.~Friot and S.~Ghosh,
Phys. Rev. D \textbf{101} (2020) no.11, 116008
doi:10.1103/PhysRevD.101.116008
[arXiv:2003.12030 [hep-ph]].

\bibitem{Ananthanarayan:2020xut}
B.~Ananthanarayan, S.~Friot, S.~Ghosh and A.~Hurier,
[arXiv:2005.07170 [hep-th]].

\bibitem{Smirnov:2012gma}
V.~A.~Smirnov,
\textit{``Analytic tools for Feynman integrals''},
Springer Tracts Mod. Phys. \textbf{250} (2012), 1-296
doi:10.1007/978-3-642-34886-0



\bibitem{Passare:1996db}
M.~Passare, A.~K.~Tsikh and A.~A.~Cheshel,
Theor. Math. Phys. \textbf{109} (1996), 1544-1555
doi:10.1007/BF02073871
[arXiv:hep-th/9609215 [hep-th]].

\bibitem{TZ}
O. Zhdanov and A. Tsikh, Siberian Mathematical Journal {\bf 39}, 245 (1998).



\bibitem{Erdelyi}

A. Erdelyi, W. Magnus, F. Oberhettinger and F. G. Tricomi, \textit{``Higher transcendental functions,''} Bateman Project Vol. 1, McGraw-Hill Book Company (1953).







  
\bibitem{Larsen:2017aqb}
K.~J.~Larsen and R.~Rietkerk,
Comput. Phys. Commun. \textbf{222} (2018), 250-262
doi:10.1016/j.cpc.2017.08.025
[arXiv:1701.01040 [hep-th]].
  


  
 

 
 
  


  
\end{thebibliography}
\end{document}